\documentclass{jfm}
\RequirePackage{expl3}
\ExplSyntaxOn
\cs_if_exist:NF \prop_gput_if_not_in:NnV
  {
    \cs_new_protected:Npn \prop_gput_if_not_in:NnV #1#2#3
      {
        \prop_if_in:NnTF #1 {#2}
          { }
          { \prop_gput:NnV #1 {#2} #3 }
      }
  }
\cs_if_exist:NF \prop_gput_if_not_in:Nnn
  {
    \cs_new_protected:Npn \prop_gput_if_not_in:Nnn #1#2#3
      {
        \prop_if_in:NnTF #1 {#2}
          { }
          { \prop_gput:Nnn #1 {#2} {#3} }
      }
  }
\ExplSyntaxOff
\makeatletter
\providecommand\IfPackageLoadedTF[3]{\@ifpackageloaded{#1}{#2}{#3}}
\providecommand\IfPackageLoadedT[2]{\@ifpackageloaded{#1}{#2}{}}
\providecommand\IfPackageLoadedF[2]{\@ifpackageloaded{#1}{}{#2}}
\makeatother
\makeatletter
\ifdefined\pdfpxdimen
\fi
\makeatother
\usepackage{mathtools}
\usepackage[T1]{fontenc}   
\usepackage{lmodern}       
\usepackage{fix-cm}
\usepackage{hyperref}
\DeclareMathSizes{10.5}{10.5}{7.35}{5.25} 
\RequirePackage[T1]{fontenc}      
\RequirePackage{newtxtext,newtxmath} 
\RequirePackage{fix-cm}              
\usepackage[section]{placeins} 
\RequirePackage[a4paper]{geometry}   
\definecolor{MyGrey}{gray}{0.4}      

\usepackage{colortbl}
\usepackage{booktabs}                

\usepackage{lineno}
\usepackage{graphicx}
\usepackage{natbib}
\usepackage{amssymb}
\usepackage{amsbsy}
\usepackage{amsmath}
\usepackage{graphics}
\usepackage{dcolumn}
\usepackage{bm}
\usepackage{hyperref}
\usepackage{xcolor}
\usepackage{soul}
\usepackage{color}
\usepackage[utf8]{inputenc}  
\usepackage[T1]{fontenc}     
\usepackage{subcaption}

\usepackage{lmodern}\usepackage{fix-cm}\DeclareMathSizes{10.5}{10.5}{7.35}{5.25}

\newcommand{\bc}{\begin{center}}
\newcommand{\ec}{\end{center}}

\newcommand{\be}{\begin{equation}}
\newcommand{\ee}{\end{equation}}
\newcommand{\bea}{\begin{eqnarray}}
\newcommand{\eea}{\end{eqnarray}}

\title[]{Fluid Deformation in Random Unsteady Flow}

\author[Daniel R. Lester and Marco Dentz]
{Daniel R. Lester$^1$%
  \thanks{Email address for correspondence: daniel.lester@rmit.edu.au},\ns
Marco Dentz$^2$}

\affiliation{$^1$School of Engineering, RMIT University, Melbourne, Australia\\[\affilskip]
$^2$Spanish National Research Council (IDAEA-CSIC), 08034 Barcelona, Spain\\[\affilskip]
}

\pubyear{2026}
\volume{}
\pagerange{}
\date{?; revised ?; accepted ?. - To be entered by editorial office}
\begin{document}

\maketitle

\begin{abstract}
Fluid deformation controls myriad processes in random flows including mixing and dispersion, stress development in complex fluids, colloid transport and deposition, droplet breakup and emulsification, fluid-structure interaction, chemical reactions and biological activity. Despite this, fundamental aspects are not well understood, including the link between the Lagrangian velocity gradient tensor $\boldsymbol\epsilon$ and deformation measures such as Lyapunov exponents ($\lambda_{\infty,i}$), their finite-time counterparts (FTLEs) and the right Cauchy-Green tensor $\mathbf{C}$. We address these knowledge gaps by developing an \emph{ab initio} stochastic model of fluid deformation in ergodic and stationary random unsteady flows. We show that although the Lagrangian velocity process is non-Markovian and non-Fickian, temporal decorrelation in unsteady random flows results in Fickian evolution of $\boldsymbol\epsilon$. Application of an objective coordinate transform renders $\boldsymbol\epsilon^\prime$ upper triangular, the basis vectors of which exponentially converge to Lyapunov vectors. As such, the diagonal components of $\boldsymbol\epsilon^\prime$ correspond to increments of the Lyapunov spectra, while the off-diagonal components objectively quantify shear and vorticity. This leads to a stochastic model of Lagrangian fluid deformation as a simple Brownian process that provides a direct link between $\boldsymbol\epsilon^\prime$ and fluid deformation. We develop closed-form expressions for the evolution of $\mathbf{C}$ and the FTLEs, and apply the stochastic  model to numerical results for a model 2D unsteady flow and 3D forced homogeneous isotropic turbulence, returning excellent agreement with direct calculations of deformation measures. This approach provides an objective means of characterising the deformation properties of unsteady flows and the development of stochastic models of deformation.
\end{abstract}


\section{Introduction}
\label{sec:intro}

Deformation of fluid elements is fundamental to myriad fluid-borne processes, ranging from stretching of material lines and surfaces~\citep{Meneveau:2011aa,Ottino:1990aa} to diffusive mixing and transport of solutes, particles and scalars~\citep{Dimotakis:2005aa,Sapsis:2010aa,Villermaux:2019aa}, fluid-structure interaction~\citep{Griffith:2020aa}, development of stresses in polymer and viscoelastic systems~\citep{Rivlin:1971aa,Wineman:2009aa}, turbulent energy cascade and dissipative structures~\citep{Yao:2022aa}, Lagrangian coherent structures (LCS) that govern advective transport~\citep{Haller:2015aa}, particle orientation, alignment and dissipation~\citep{Voth:2017aa}, multiphase processes such as droplet breakup and emulsification~\citep{Stone:1994aa}, and reactive processes including chemical reactions~\citep{Libby:1976aa}, biological activity~\citep{Tel:2005aa,Neufeld:2009aa} and geochemical processes~\citep{Lester:2012aa}. The understanding, characterisation and prediction of these processes requires quantification of fluid deformation, and indeed many models of these processes require characterisation of fluid shear and stretching rates (such the Lyapunov spectra) as model inputs.

To leading order, fluid deformation is characterised in terms of the fluid deformation gradient tensor $\mathbf{F}(\mathbf{X},t)\equiv d\mathbf{x}/d\mathbf{X}$ (or the right Cauchy-Green tensor $\mathbf{C}(\mathbf{X},t)\equiv\mathbf{F}^\top\mathbf{F}$), which quantifies the affine deformation of material elements between the Eulerian $\mathbf{x}$ and Lagrangian $\mathbf{X}$ frames, and evolves in 
Lagrangian time $t$ along pathlines of the unsteady velocity field $\mathbf{v}(\mathbf{x},t)$ as
\begin{equation}
\frac{d\mathbf{F}(\mathbf{X},t)}{dt}=\boldsymbol\epsilon(\mathbf{X},t)\mathbf{F}(\mathbf{X},t),\quad\mathbf{F}(\mathbf{X},0)=\mathbf{I},\label{eqn:deform}
\end{equation}
where $\boldsymbol\epsilon(\mathbf{X},t)\equiv \nabla\mathbf{v}[\mathbf{x}(\mathbf{X},t),t]^\top$ is the Lagrangian velocity gradient tensor. For random unsteady flows such as turbulent flows~\citep{Girimaji:1990aa}, sheared suspensions~\citep{Souzy:2017aa} and transient flows in random porous media~\citep{Tartakovsky:2004aa}, the prediction of fluid deformation dates back to consideration of line and surface stretching by \citep{Batchelor:1952aa}, followed by a series of studies~\citep{Cocke:1969aa,Orszag:1970aa,Cocke:1971aa,Lin:1973aa,Girimaji:1990aa,Dresselhaus:1992aa,Tabor:1994aa,Villermaux:1994aa,Kalda:2000aa,Goto:2002aa,Villermaux:2019aa} that examine the deformation kinematics of material elements (lines and surfaces) from a stochastic perspective. The majority of these studies involve phenomenological models of deformation that assume deformation dynamics; lacking is a rigorous derivation  from first principles.

Furthermore, the link between relevant deformation characteristics such as Lyapunov exponents $\lambda_{\infty,i}$ (with $\lambda_{\infty,i}\geqslant\lambda_{\infty,i+1}$ and $i=1:d$ for $d$-dimensional flows) and their finite-time variants (FTLEs) 
is unclear as the principal directions of strain (i.e., that correspond to the eigenvectors $\mathbf{c}_i$ of $\mathbf{C}$) evolve due to both rotation (via the vorticity tensor $\boldsymbol\Omega(\mathbf{X},t)\equiv\small{\frac{1}{2}}(\boldsymbol\epsilon-\boldsymbol\epsilon^\top)$) and strain (via the strain rate tensor $\mathbf{S}(\mathbf{X},t)\equiv\small{\frac{1}{2}}(\boldsymbol\epsilon+\boldsymbol\epsilon^\top)$) of fluid elements. Typically, Lyapunov exponents are determined by first computing fluid deformation and subsequently computing these metrics~\citep{Greene:1987aa,Cui:2021aa}, which can lead to numerical instabilities~\citep{Dieci:1997aa}. In the same way that FTLEs converge with time to $\lambda_{\infty,i}$, the eigenvectors $\mathbf{c}_i$ converge to the covariant Lyapunov vectors $\mathbf{a}_i$~\citep{Ginelli:2013aa} (otherwise known as Oseledec invariant directions~\citep{Oseledec:1968aa,Ruelle:1979aa}) that are stretching directions associated with Lyapunov exponents $\lambda_{\infty,i}$.

This evolution is precisely why Eulerian measures fail to properly characterise fluid deformation or, e.g., correctly detect coherent structures such as LCS~\citep{Haller:2015aa} (which are ultimately demarcated by their local deformation properties). Deformation is an inherently Lagrangian process that requires the evolving material stretching directions to be resolved.  As a result, averaging of the eigenvalues of the Lagrangian strain rate tensor $\mathbf{S}(\mathbf{X},t)$ (which account for material rotation by implicitly rotating the frame of reference with respect to the vorticity tensor $\boldsymbol\Omega(\mathbf{X},t)$) do not correspond to the infinite-time Lyapunov exponents $\lambda_{\infty,i}$~\citep{Ottino:1989aa}. This raises the question -  is there a representation of $\boldsymbol\epsilon$ that corresponds to the covariant Lyapunov vectors $\mathbf{a}_i$? Such a representation would provide a direct link between $\boldsymbol\epsilon$ and deformation measures such as FTLEs, $\lambda_{\infty,i}$ and $\mathbf{C}$ as well as provide a basis for \emph{ab inito} models of fluid deformation.

For random steady flows, a moving and rotating coordinate system termed the \emph{Protean} frame~\citep{Adachi:1983aa}, defined as that which renders the rotated velocity gradient $\boldsymbol{\epsilon}^\prime$ upper triangular, has previously been employed to develop accurate stochastic models of fluid deformation in random steady 2D~\citep{Dentz:2016aa} and 3D~\citep{Lester:2018aa} flows, such as those arising in heterogeneous porous media. In steady 2D flows, the Protean transformation reduces to a moving streamline coordinate system. This representation is essential for constructing stochastic models that capture the inherently algebraic-in-time deformation characteristic of integrable motion~\citep{Dentz:2016aa}, reflecting the topological constraints imposed by steady planar flow. In 3D steady flows, the Protean frame likewise aligns with streamlines, but additionally permits exponential stretching associated with chaotic advection~\citep{Lester:2018aa}, but only if the helicity density $\mathbf{h}\equiv\mathbf{v}\cdot(\nabla\times\mathbf{v})$~\citep{Moffatt:1969aa} of the steady 3D flow is non-zero~\citep{Lester:2022aa}, otherwise the flow is \emph{epi-2D}~\citep{Yoshida:2017aa}.  

As the Protean frame represents a continuous QR decomposition of $\boldsymbol{\epsilon}$, it shares many parallels with dynamical systems methods~\citep{Dieci:1997aa} for estimation of Lyapunov spectra from the diagonal components of $\boldsymbol{\epsilon}^\prime$. As demonstrated above, the Protean frame also enforces topological constraints in non-chaotic systems, where the off-diagonal components of $\boldsymbol{\epsilon}^\prime$ control deformation. In Appendix~\ref{app:strip} we also show that the Protean frame is equivalent to rotation into a coordinate system aligned with a 1D infinitesimal material strip $\mathbf{l}(t)$ that is used in many fluid mixing models~\citep{Villermaux:2019aa}, and the growth of $|\mathbf{l}(t)|$ is governed by $\epsilon^\prime_{11}$. In general, the deformation of 2D and 3D material objects are governed by all the non-zero elements of $\boldsymbol{\epsilon}^\prime$. Although the Protean frame naturally recovers the deformation characteristics and enforces inherent topological constraints in a wide range of flow classes, it has not been previously applied to unsteady random flows.

A further issue is the Lagrangian evolution of $\boldsymbol\epsilon$ along trajectories. The Lagrangian velocity process in random, unsteady flow such as homogeneous isotropic turbulence (HIT) is non-Markovian in space and time, leading to strong intermittency of $\boldsymbol\epsilon$~\citep{Meneveau:2011aa}. Furthermore, transport in many unsteady flows such as HIT or flow in highly heterogeneous porous media are also strongly non-Fickian~\citep{Brandenburg:2004aa,Levy:2003aa,Bijeljic:2011aa,Comolli2019,Puyguiraud2019}, leading to complex evolution of Lagrangian velocity~\citep{Le-Borgne:2008aa,Le-Borgne:2008ab,Hakoun2019} and velocity gradients~\citep{Dentz:2016aa,Lester:2018aa}. Hence it is unclear what is an appropriate stochastic framework for evolution of $\boldsymbol\epsilon$~\citep{Meneveau:2011aa}.

In this study we address these challenges by developing an \emph{ab initio} stochastic model of fluid deformation in  ergodic and statistically stationary random unsteady flows. This stochastic deformation model is based upon exponential decay of the temporal Lagrangian velocity gradient autocorrelation function, leading to evolution of the components of the velocity gradient components which are well-described via Brownian motion. We employ the Protean frame to render $\boldsymbol\epsilon^\prime$ upper-triangular and objective, and show that this coordinate frame rapidly converges to the covariant Lyapunov vectors $\mathbf{a}_i$. As the Protean frame also renders the deformation tensor upper triangular, analytic solutions are available for $\mathbf{F}^prime$, facilitating the development of stochastic models for its evolution and the Cauchy-Green tensor and associated FTLEs. We apply this method to numerical data for a model 2D unsteady flow and direct numerical simulation (DNS) simulations of forced HIT to test how well this model characterises all aspects of fluid deformation evolution. Hence this method represents a useful method for statistical characterisation of velocity gradients and fluid deformation in random unsteady flows facilitating identification of Lyapunov spectra and stochastic prediction of deformation evolution.

The remainder of this paper is structured as follows. In \S~\ref{sec:decorrelation} we consider the temporal evolution of the Lagrangian velocity gradient tensor in random unsteady flow and develop a simple stochastic model for this process, along with objectivity of deformation measures. These results are used in \S~\ref{sec:deform} to develop a coordinate transform for $\boldsymbol\epsilon$ that rapidly converges to the Lyapunov vectors and ultimately develop a stochastic model for the deformation evolution. To test this model and demonstrate application, in \S~\ref{sec:numerics} this model is then applied to numerical data for a model 2D flow and DNS data of HIT. Conclusions and research directions are discussed in \S~\ref{sec:conclusions}.

\section{Velocity Gradients and Deformation in Random Flows}
\label{sec:decorrelation}

\subsection{Non-Fickian Transport}
\label{subsec:nonfickian}

An important consideration is the development of an appropriate stochastic framework for evolution of $\boldsymbol\epsilon$ in random unsteady flow. For random steady flows, the Lagrangian velocity decorrelates in space as tracer particles are advected through the spatially heterogeneous velocity field~\citep{Le-Borgne:2008aa,Le-Borgne:2008ab}, and hence follow a spatial Markov process, leading to  continuous time or time-domain random walk models for the evolution of longitudinal dispersion in heterogeneous porous media~\citep{Berkowitz:2006aa,Dentz:2016ab,Noetinger2016}. Such spatial Markovianity leads to non-Fickian transport if the Eulerian velocity distribution $p_e(v)$ for $v\ll \langle v\rangle$ scales as $p_e(v)\sim v^{\beta-1}$ with $0<\beta<1$. Further studies~\citep{Dentz:2016aa,Lester:2018aa} have shown that, as expected, the velocity gradient tensor component in steady random flows has the same decorrelation structure as the velocity vector components, and so also follows a spatial Markov process, leading to CTRW models for fluid deformation in random steady 2D~\citep{Dentz:2016aa} and 3D~\citep{Lester:2018aa} flows which also exhibit non-Fickian behaviour if $p_e(v)\sim v^\beta$ for $v\ll \langle v\rangle$ with $1<\beta<2$.

\subsection{Non-Markovian Dynamics}
\label{subsec:nonmarkovian}

For unsteady flows such as turbulent flows or transiently forced flows in heterogeneous porous media, the picture is more complicated as these flows are observed to be neither Markovian in space or time, leading to strong intermittency of both the Lagrangian velocity and velocity gradient in time. However, if these flows are space-time ergodic (i.e., the statistics are sampled by the space and time fluctuations), these quantities may be rendered Markovian with respect to a Lagrangian sampling variable $r(s,t)$~\citep{Dentz:2025aa} that is a function of the distance travelled $s$ and time $t$ along a pathline, which is chosen such that the sequence of Lagrangian particle speeds is Markovian with respect to $r$.

This leads to the notion of a local Kubo number $\kappa_n=v_n \tau_c/\ell_v$ (where $v_n$ is the local velocity and $\tau_c$, $\ell_v$ respectively are the characteristic decorrelation time and length) that characterises the competition between local spatial and temporal decorrelation. For $\kappa_n\ll 1$, the local velocity $v_n$ decorrelates in time, whereas for $\kappa_n\gg 1$, $v_n$ decorrelates in space, and for $\kappa\sim 1$, velocity decorrelation is spatio-temporal. This formulation renders the velocity process Markovian with respect to $r$, hence a CTRW model can be developed that captures the evolution of the Lagrangian velocity in $r$-space.

For space-time non-separable unsteady flows (i.e., those whose velocity field cannot be decomposed into a spatial field multiplied by a transient forcing function), transport is always Fickian (even if $0<\beta<1$) on scales longer than the temporal decorrelation scale $\tau_c$~\citep{Dentz:2025aa}, as long episodes in low velocity regions (that would normally generate persistent non-Fickian behaviour in steady or space-time separable flows) are interrupted by resetting of the temporal velocity process. Hence, despite the non-Markovian properties of ergodic unsteady flows, the evolution of the Lagrangian velocity (and hence velocity gradient) in time is remarkably simple and is described by a Brownian process along pathlines.


\subsection{Deformation and Objectivity}
\label{subsec:objective}

A key property for correct development of constitutive laws~\citep{Wineman:2009aa} and identification of kinematic structures such as vortices~\citep{Haller:2005aa,Haller:2015aa} is that physical quantities are \emph{objective}~\citep{Truesdell:1992aa,Tadmor:2011aa} in that they are \emph{frame-indifferent} and so represent physical quantities which are independent of the rotating and translating rigid frame of reference
\begin{equation}
\mathbf{x}^\prime = \mathbf{b}(t)+\mathbf{Q}^\top(t)\mathbf{x},\label{eqn:transform}
\end{equation}
where $\mathbf{Q}(t)$ is a proper orthogonal rotation matrix that satisfies $\mathbf{Q}^\top\mathbf{Q}=\mathbf{I}$ and $\mathbf{Q}(0)=\mathbf{I}$, and $\mathbf{c}(t)$ encodes translation. Conversely, Galelian invariance, which requires $\mathbf{b}(t)=\mathbf{a}\,t$, has been shown~\citep{Haller:2005aa} to incorrectly identify coherent structures.

Under fluid deformation we differentiate between \emph{spatial} measurements in the current (Eulerian) configuration $\mathbf{x}$ and \emph{material} measurements in the reference (Lagrangian) configuration $\mathbf{X}$, and \emph{two-point} tensors such as $\mathbf{F}$, which are associated with both configurations. Under this classification, arbitrary scalars $\phi$, spatial vectors $\mathbf{a}$ and spatial second order tensors $\mathbf{T}$ are considered objective if they transform under (\ref{eqn:transform}) according to
\begin{align}
\phi^\prime=\phi && \mathbf{a}^\prime=\mathbf{Q}^\top\mathbf{a} && \mathbf{T}^\prime=\mathbf{Q}^\top\mathbf{T} \mathbf{Q},\label{eqn:spatial_objectivity}
\end{align}
such that these quantities are \emph{true} vectors and tensors (as opposed to pseudovectors and pseudotensors \citep{Ottino:1989ab}). Similarly, material vectors and tensors are objective if they transform as $\mathbf{a}^\prime=\mathbf{a}$ and $\mathbf{T}^\prime=\mathbf{T}$, and two-point tensors $\mathbf{T}$ are objective if they transform as $\mathbf{T}^\prime=\mathbf{Q}^T\mathbf{T}\,$ \footnote{Note that some authors do not differentiate between spatial and material tensors and so only consider second order tensors to be objective with respect to (\ref{eqn:spatial_objectivity}).}. 
While the spatial tensor $\boldsymbol\epsilon$ is generally non-objective (and so its invariants $P,Q,R$ must be interpreted with caution as they are not frame-indifferent), the strain-rate spatial tensor $\mathbf{S}$ and vorticity spatial tensor $\boldsymbol\Omega$ are objective~\citep{Tadmor:2011aa}, as are their scalar invariants such as shear rate $\dot\gamma\equiv\sqrt{2\text{tr}(\mathbf{S}^2)}$ and vorticity $\omega\equiv\sqrt{2\text{tr}(\boldsymbol\Omega^2)}$. Similarly, the deformation gradient two-point tensor $\mathbf{F}$, the left Cauchy-Green spatial tensor $\mathbf{B}\equiv\mathbf{F}\mathbf{F}^\top$ and the right Cauchy-Green material tensor $\mathbf{C}^\prime=\mathbf{C}$ and associated FTLEs and Lyapunov exponents are also all objective:
\begin{equation}
    \lambda_i(\mathbf{X},t;t_0)\equiv\frac{1}{2|t-t_0|}\ln\sigma_i(\mathbf{X},t),\quad \lambda_{\infty,i}\equiv\lim_{t\rightarrow\infty}\lambda_i(\mathbf{X},t),\label{eqn:FTLE1}
\end{equation}
where $\sigma_i(\mathbf{X},t)$ are the eigenvalues of $\mathbf{C}$. In the following, objective transforms and measures will be used to develop stochastic models of fluid deformation.

\subsection{Lyapunov Vectors and Exponents}
\label{subsec:lyapunov}

As such we seek an objective representation of the velocity gradient tensor $\boldsymbol\epsilon$ that naturally encodes this process. Such a representation can be obtained by generalising (\ref{eqn:transform}) to a materially-dependent transform along a tracer particle trajectory where $\mathbf{b}(t)\mapsto\mathbf{x}_0(\mathbf{X},t)$, $\mathbf{Q}(t)\mapsto\mathbf{Q}(\mathbf{X},t)$ and $\mathbf{x}_0(\mathbf{X},t=0)\equiv\mathbf{X}$ is the initial location of the tracer. Although this transform does not correspond to a rigid body rotation globally (due to $\mathbf{X}$-dependence of $\mathbf{Q}$, $\mathbf{x}_0$), it does correspond to a rigid frame along a single particle trajectory and so maintains objectivity in this sense. Due to ergodicity, this transform can also be applied to multiple distinct trajectories for the purposes of gathering statistics to objectively characterise fluid deformation. Under this transform, the Lagrangian velocity gradient (spatial) tensor $\boldsymbol\epsilon$ transforms as a pseudotensor~\citep{Truesdell:1992aa,Lester:2018aa}
\begin{equation}
\boldsymbol\epsilon^\prime(\mathbf{X},t)=\mathbf{Q}(\mathbf{X},t)^\top\boldsymbol\epsilon(\mathbf{X},t)\mathbf{Q}(\mathbf{X},t)+\frac{d\mathbf{Q}(\mathbf{X},t)^\top}{dt}\mathbf{Q}(\mathbf{X},t),\label{eqn:rotn}
\end{equation}
and so is not objective. Note that the deformation evolution equation (\ref{eqn:deform}) (with primes added) also holds in this moving coordinate frame.

Objectivity alone is not sufficient to properly characterise fluid deformation. As material elements are rotated by both vorticity $\boldsymbol\Omega$ and strain $\mathbf{S}$, Lyapunov exponents cannot be determined from Eulerian measures such as $\mathbf{S}$. A physical picture of deformation is obtained by consideration of the deformation of an infinitesimal sphere of radius $\delta r$ at initial time $t=t_0$, such that $||d\mathbf{X}||^2=d\mathbf{X}^\top d\mathbf{X}=\delta r^2$, then using $d\mathbf{x}=\mathbf{F}d\mathbf{X}$, the current configuration at time $t$ is $||d\mathbf{x}||^2=d\mathbf{x}^\top d\mathbf{x}=\delta r^2$ with $\sigma_i(t_0)=1$. As $\mathbf{C}$ is diagonalizable, then $\mathbf{C}=\mathbf{R}\boldsymbol\Sigma\mathbf{R}^\top$ where $\mathbf{R}$ is a rotation tensor and $\boldsymbol\Sigma=\text{diag}(\sigma_1(t),\sigma_2(t),\sigma_3(t))$, then defining $\mathbf{z}\equiv\mathbf{R}^\top d\mathbf{X}$, so $z_i=\delta r$ and $||d\mathbf{x}||^2=\sum_{i=1}^3\sigma_i(t) z_i^2$. Hence the infinitesimal sphere deforms with time $t>t_0$ as an ellipsoid with semi-axes of length $\sqrt{\sigma_i(t)}\delta r$, where the eigenvalues $\sigma_i(t)$ of $\mathbf{C}$ grow exponentially in time as
\begin{equation}
\dot\sigma_i(t)=2\sigma_i(t)\mathbf{F}^\top\mathbf{S}\mathbf{F},
\end{equation}
and the ellipsoid semi-axes are given by the eigenvectors $\mathbf{c}_i$ of $\mathbf{C}$. These eigenvectors converge as $1/\sqrt{t-t_0}$ to the covariant Lyapunov vectors $\mathbf{a}_i$.
%

Under the Protean coordinate frame, an appropriate choice of rotation $\mathbf{Q}(\mathbf{X},t)$ along particle trajectories renders both $\boldsymbol\epsilon^\prime(\mathbf{X},t)$ and $\mathbf{F}^\prime(\mathbf{X},t)$ (from (\ref{eqn:deform})) upper triangular, yielding simple closed-form expressions for the components of $\mathbf{F}^\prime(\mathbf{X},t)$ in terms of the elements of $\boldsymbol\epsilon^\prime(\mathbf{X},t)$. For steady flows this frame corresponds to a streamline coordinate system (where $\hat{\mathbf{e}}_1^\prime=\mathbf{v}/v$), whereas for unsteady flows, the basis vectors $\hat{\mathbf{e}}_1^\prime$ converge exponentially with time $t-t_0$ to both the covariant Lyapunov vectors $\mathbf{a}_i$.

\subsection{Continuous QR Decomposition}
\label{subsec:QR}

Transform of $\boldsymbol\epsilon$ into the Protean frame is equivalent to application of a continuous QR decomposition of $\mathbf{F}^\prime$ along particle trajectories which forms the basis of methods~\citep{Dieci:1997aa,Lu:2005aa,Froyland:2013aa} to estimate Lyapunov spectra in dynamical systems. Although these QR methods focus on Lyapunov vectors and exponents, the Protean frame also correctly enforces topological constraints in non-chaotic flows~\citep{Dentz:2016aa,Lester:2018aa} and, as shall be shown, the shear and vorticity (off-diagonal) components of $\boldsymbol\epsilon^\prime$ also play an important role in controlling deformation. For bounded and continuous $\boldsymbol\epsilon(\mathbf{X},t)$, this QR decomposition exists and is unique~\citep{Dieci:1997aa}, and the Lyapunov exponents and spectra of $\mathbf{F}^\prime$ are the same as that of $\mathbf{F}$~\citep{Dieci:1997aa}. As $\dot{\mathbf{Q}}(\mathbf{X},t)^\top\mathbf{Q}(\mathbf{X},t)$ in (\ref{eqn:rotn}) is skew-symmetric, then the diagonal elements of $\mathbf{F}^\prime$ evolve as
\begin{equation}
\dot{F}_{ii}^\prime=\tilde\epsilon_{ii}F_{ii}^\prime,\quad i=1:d,
\end{equation}
where $\tilde{\boldsymbol\epsilon}\equiv\mathbf{Q}^\top\boldsymbol\epsilon\mathbf{Q}$ and so the Lyapunov exponents are given as temporal averages of $\tilde\epsilon_{ii}=\epsilon^\prime_{ii}$ as
\begin{equation}
\lambda_{\infty,i}=\lim_{t\rightarrow\infty}\frac{1}{t}\ln F^\prime_{ii}(t)=\lim_{t\rightarrow\infty}\frac{1}{t}\int_0^t \tilde\epsilon_{ii}(s)ds=\langle\tilde\epsilon_{ii}\rangle=\langle\epsilon^\prime_{ii}\rangle,\quad i=1:d,\label{eqn:LyapunovQR}
\end{equation}
and we denote the maximum Lyapunov exponent as $\lambda_\infty=\lambda_{\infty,1}$. For all dynamical systems, the existence of distinct Lyapunov exponents $\lambda_{\infty,i}$, and associated invariant directions $\mathbf{a}_i$ (linear subspaces) is given by Oseledec's theorem~\citep{Oseledec:1968aa,Ruelle:1979aa}. Several authours~\citep{Froyland:2013aa,Noethen:2019aa} have shown that the QR subspaces converge exponentially in time to the Oseledec subspaces at a rate proportional to the spectral gap between Lyapunov exponents as
\begin{equation}
\theta_i(t)\sim\exp\left[(\lambda_{\infty,i+1}-\lambda_{\infty,i})t\right],\label{eqn:spectral_gap}
\end{equation}
where $\theta_i(t)$ is the angle between the QR basis vectors $\hat{\mathbf{e}}^\prime_i$ and the covariant Lyapunov vectors $\mathbf{a}_i$. Hence both the Protean frame and the continuous QR decomposition yields accurate representation of the covariant Lyapunov vectors for advection times $t-t_0$ significantly larger than the \emph{spectral gap time} $\tau_{\Delta\lambda}\equiv1/\min_i|\lambda_{\infty,i+1}-\lambda_{\infty,i}|$. We also note that for non-chaotic flows, the absence of non-zero Lyapunov exponents means that the QR transform is explicit and aligns exactly with the flow invariants (such as streamlines) at all times~\citep{Dentz:2016aa,Lester:2021aa}.

Although the velocity gradient tensor $\boldsymbol\epsilon$ is in general not objective, both the diagonal and off-diagonal components of the Protean velocity gradient tensor $\boldsymbol\epsilon^\prime$ are all objective as the continuous QR decomposition is unique for $t\gg\tau_{\Delta\lambda}$. Hence the elements of $\boldsymbol\epsilon^\prime$ are all independent of any objective change of frame prior to application of the Protean transform. The objective nature of $\boldsymbol\epsilon^\prime$ is key to correctly quantifying deformation in arbitrary flow fields.

\section{Fluid Deformation in the Protean Frame}
\label{sec:deform}

\subsection{Rotation and Deformation in the Protean Frame}
\label{subsec:protean}

\subsubsection{Two-dimensional unsteady flow} To illustrate rotation into the Protean frame for simplicity we first consider composition of $\mathbf{Q}(\mathbf{X},t)$ in 2D planar unsteady flows, where rotation of angle $\alpha_3(\mathbf{X},t)$ about the $x_3$ coordinate normal to the $x_1-x_2$ plane containing the flow is given by $\mathbf{Q}(\mathbf{X},t)=\mathbf{Q}_3(\alpha_3(\mathbf{X},t))$, where
\begin{equation}
\begin{split}
\mathbf{Q}_i(\alpha_i)&=\cos\alpha_i\,\mathbf{I}+\sin\alpha_i(\hat{\mathbf{e}}_i)_\times+(1-\cos\alpha_i)\,\hat{\mathbf{e}}_i\otimes\hat{\mathbf{e}}_i,,\label{eqn:Qi}
\end{split}
\end{equation}
and $(\,)_\times$ denotes the cross-product matrix
\begin{equation}
(\mathbf{a})_\times\equiv
\left(\begin{array}{ccc}
0 & -a_3 & a_2\\
a_3 & 0 & -a_1\\
-a_2 & a_1 & 0
\end{array}
\right).\label{eqn:cross}
\end{equation}
In general, this representation of $\mathbf{Q}$ via $\alpha_i$, $i=1:d(d-1)/2$ circumvents numerical issues regarding maintenance of orthogonality of $\mathbf{Q}$~\citep{Dieci:1997aa} and reduces the number of unknowns of $\mathbf{Q}$ from $d^2$ to $d(d-1)/2$~\citep{Udwadia:2001aa}. For 2D flows with $d=2$ this means that $\mathbf{Q}$ is characterised by $d(d-1)/2=1$ angle $\alpha_3(\mathbf{X},t)$ which, via insertion of (\ref{eqn:Qi}) into (\ref{eqn:rotn}) and setting $\epsilon^\prime_{23}=0$, evolves according to
\begin{equation}
    \frac{\partial\alpha_3}{\partial t}=g[\alpha_3(t),t]\equiv
    \epsilon_{21}\cos^2\alpha_{3}-
    \epsilon_{12}\sin^2\alpha_{3}-
    \epsilon_{11}\sin(2\alpha_{3}),\quad\alpha_3(t=0)=\alpha_{3,0},\label{eqn:alpha3}
\end{equation}
where $\mathbf{X}$ has been suppressed for simplicity and $\epsilon_{22}(t)=-\epsilon_{11}(t)$ due to the divergence-free condition. Similar to the rotation angle about streamlines for the Protean frame in steady 3D flows~\citep{Lester:2018aa}, this ODE has non-zero divergence as
\begin{equation}
    \frac{\partial g}{\partial\alpha_3}=-2\cos(2\alpha_3)\sin(\alpha_3)\epsilon_{11}-2\cos(\alpha_3)\sin(\alpha_3)(\epsilon_{11}+\epsilon_{12}),
\end{equation}
with maxima and minima at $\alpha_{3}^\pm=\arctan(b\pm\sqrt{1+b^2})$, with $b\equiv (\epsilon_{12}+\epsilon_{21})/\epsilon_{11}$ where $\partial g/\partial \alpha_3>0$ at $\alpha_3=\alpha_3^+$ and $\partial g/\partial \alpha_3<0$ at $\alpha_3=\alpha_3^-$. Hence the ODE (\ref{eqn:alpha3}) has attracting ($\alpha_3^-$) and repelling ($\alpha_3^+$) states, and so admits an \emph{inertial manifold}~\citep{Guckenheimer:2002aa} $\mathcal{M}(t)$ that attracts solutions with different initial conditions $\alpha_{3,0}$, as is shown in Figure~\ref{fig:Qconverge}a. As $\epsilon_{ij}(t)$ all evolve with time, the attracting manifold $\mathcal{M}(t)$ does not correspond to $\alpha_{3}^-(t)$ (unless $\epsilon_{ij}(t)$ evolve slowly with respect to the dissipation rate of $g$ local to $\alpha_3^-(t)$) but instead ``shadows'' $\alpha_3(t)$ in a similar manner as the rotation angle in steady 3D flow~\citep{Lester:2018aa}. This attracting manifold corresponds to convergence of the Protean basis vectors $\hat{\mathbf{e}}^\prime_i$ to the covariant Lyapunov vectors $\mathbf{a}_i$ given by (\ref{eqn:spectral_gap}), hence convergence of solutions of (\ref{eqn:alpha3}) for arbitrary initial conditions $\alpha_{3,0}$ are expected to converge to $\mathcal{M}(t)$ exponentially with the spectral gap time as $\exp(-t/\tau_{\delta\lambda})$.

\subsubsection{Three-dimensional unsteady flow} For unsteady 3D flows ($d=3$) the situation is more complicated as there exist $d(d-1)/2=3$ independent angles that govern $\mathbf{Q}$. As such. we compose $\mathbf{Q}(\mathbf{X},t)$ in 3D unsteady flows as a series of rotations of angle $\alpha_1(t)$, $\alpha_2(t)$, $\alpha_3(t)$ respectively about the $x_1$, $x_2$, $x_3$ coordinates as $\mathbf{Q}(\mathbf{X},t)=\mathbf{Q}_3(\alpha_3(\mathbf{X},t))\mathbf{Q}_2(\alpha_2(\mathbf{X},t))\mathbf{Q}_1(\alpha_1(\mathbf{X},t))$, where $\mathbf{Q}_i(\alpha_i)$ are given by (\ref{eqn:Qi}).
%
%
%
%
Insertion of (\ref{eqn:Qi}) and (\ref{eqn:cross}) into (\ref{eqn:rotn}) to yield $\epsilon^\prime_{ij}(t)=0$ for $i<j$ generates a series of three coupled ODEs for the angles $\alpha_i(\mathbf{X},t)$,
\begin{align}
\frac{\partial\alpha_i(\mathbf{X},t)}{\partial t}=g_i[\alpha_1,\alpha_2,\alpha_3,\boldsymbol\epsilon],\quad\alpha_i(0)=\alpha_{i,0},\quad i=1:3,\label{eqn:rotns}
\end{align}
%
\begin{figure}
\centering
\begin{tabular}{c c c}
\includegraphics[height=0.3\columnwidth]{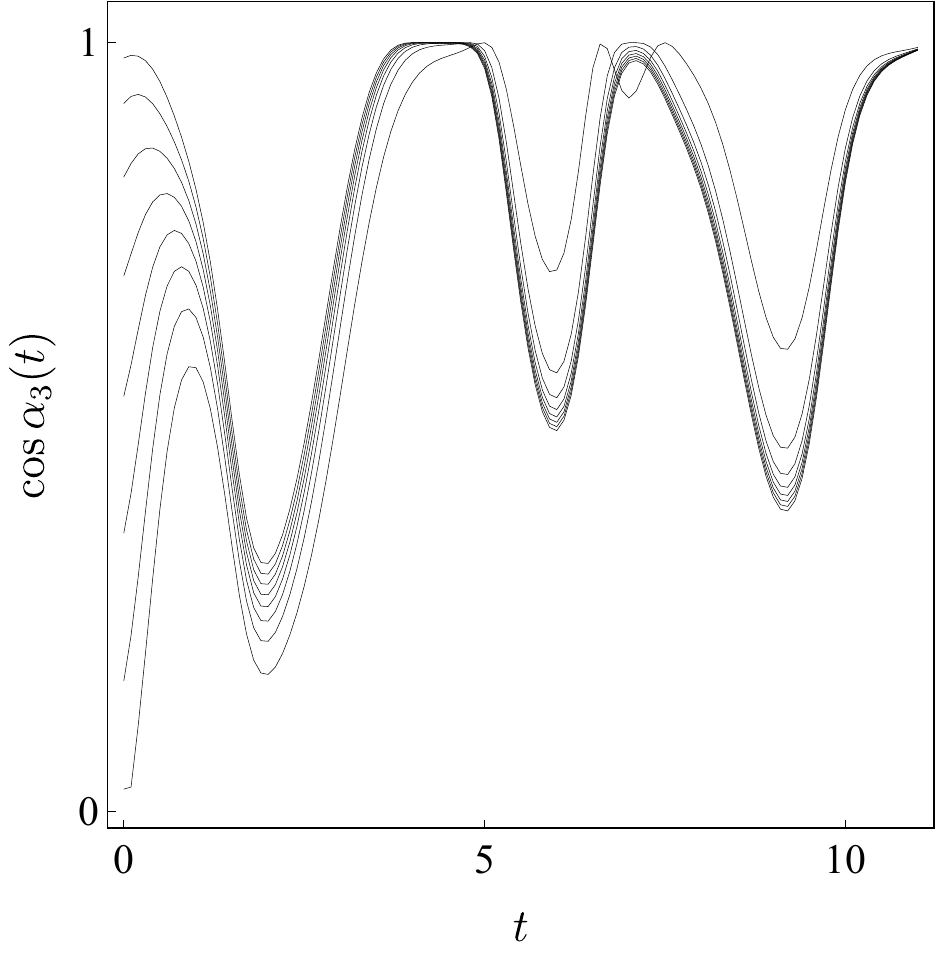}&
\includegraphics[height=0.3\columnwidth]{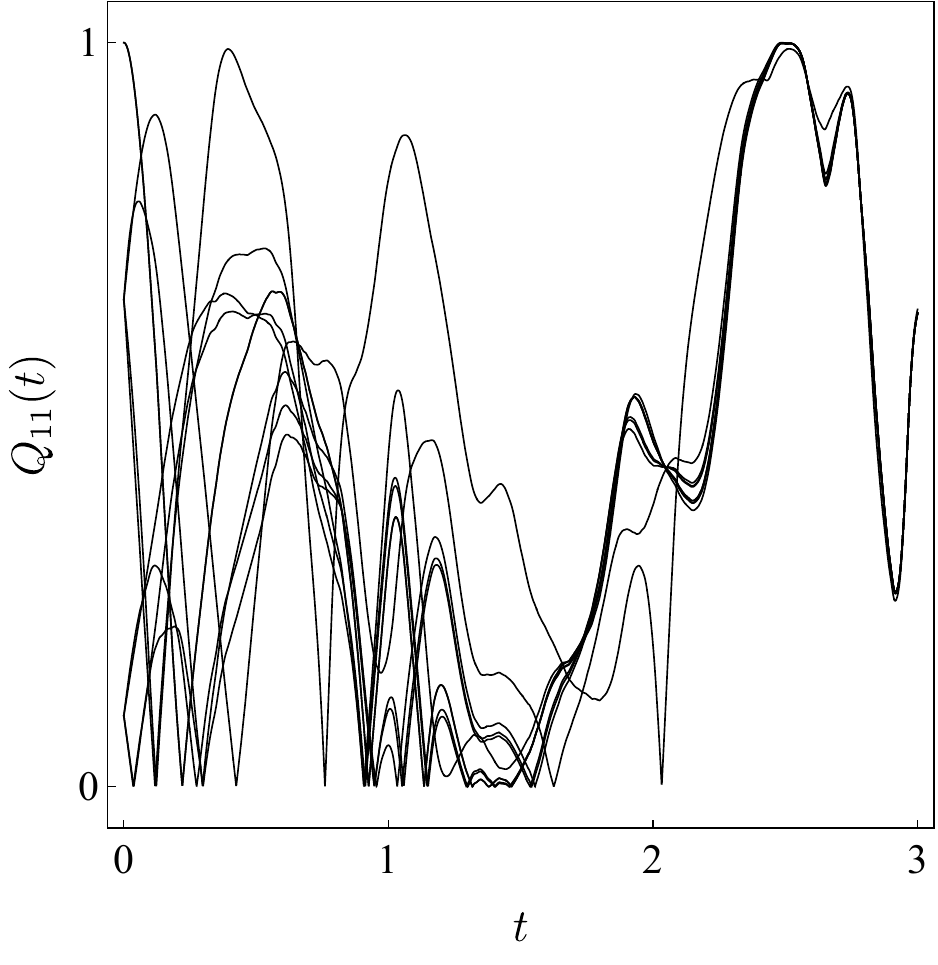}&
\includegraphics[height=0.3\columnwidth]{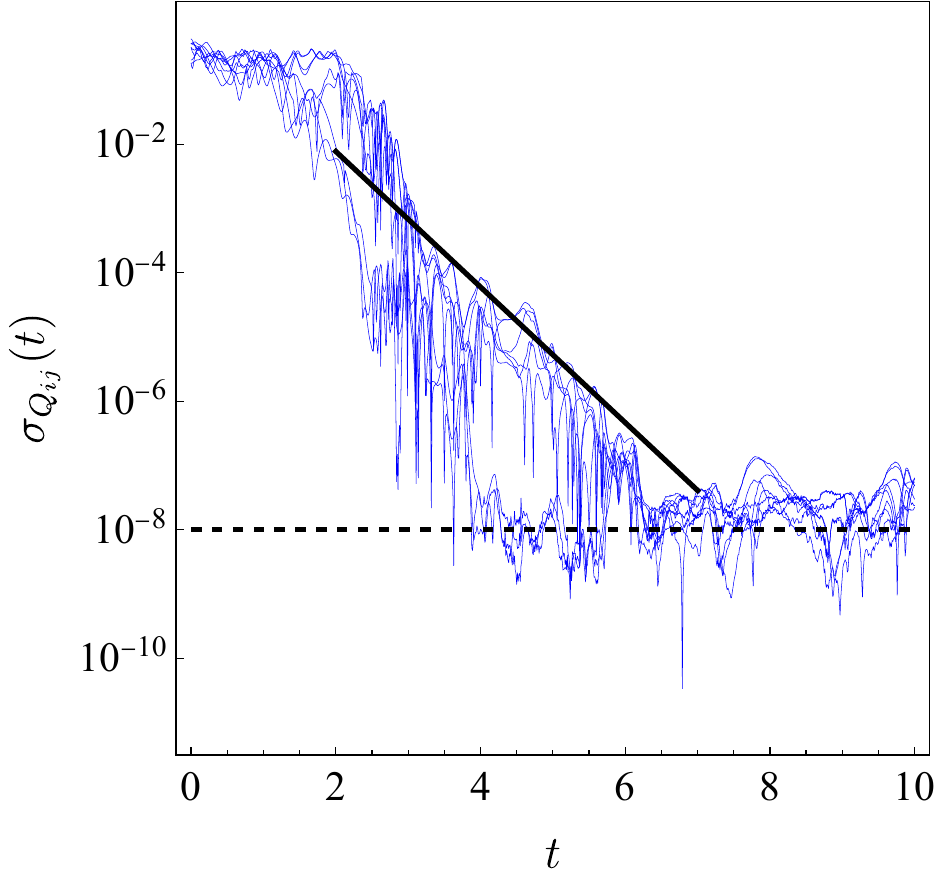}\\
(a) & (b) & (c)
\end{tabular}
\caption{ (a) Convergence of $\cos\alpha_3(t)$ with time to the attracting manifold $\mathcal{M}(t)$ for eight different initial orientation angles $\alpha_3(t=0)$ for the 2D unsteady random Kraichnan flow considered in \S~\ref{sec:numerics}. (b) Convergence of $Q_{11}(t)$ component of $\mathbf{Q}(t)$ with time for ten different combinations of initial orientation angles $\alpha_1(t=0)$, $\alpha_2(t=0)$, $\alpha_3(t=0)$, with $\alpha_i(0)\in[-\pi,\pi]$ for a typical particle trajectory in the 3D unsteady HIT flow considered in \S~\ref{sec:numerics}. Similar behaviour is observed for all other components $Q_{ij}(t)$. The converged trajectory $Q_{11}(t)$ represents an attractive inertial manifold $\mathcal{M}(t)$ of the ODE system (\ref{eqn:rotns}) that corresponds to the covariant Lyapunov vectors of the flow. (c) Convergence of the standard deviation $\sigma_{Q_{ij}}(t)$ (blue lines) between the different initial conditions $\alpha_i(t=0)$ of the components $Q_{ij}(t)$ toward the integration precision (black dashed line). The standard deviation appears to converge at a rate given by the minimum spectral gap (solid black line) $\exp(-t/\tau_{\Delta\lambda})$.}\label{fig:Qconverge}
\end{figure}
%
where the explicit form of $g_i$ is given in Appendix~\ref{app:alpha}. Hence solution of (\ref{eqn:rotns}) renders the Protean velocity gradient tensor $\boldsymbol\epsilon^\prime(\mathbf{X},t)$ upper triangular. 
Similar to the 2D case, the ODE system (\ref{eqn:rotns}) also has non-zero dissipation $\partial g_i/\partial\alpha_i\neq 0$  with an attractive branch $\partial g_i/\partial\alpha_i < 0$, hence the system (\ref{eqn:rotns}) also admits an inertial manifold $\mathcal{M}(t)$ that corresponds to the \emph{flag manifold} of $\mathbf{Q}(t)$. Hence the elements of $\mathbf{Q}(\mathbf{X},t)=\mathbf{Q}_3(\alpha_3(\mathbf{X},t))\,\mathbf{Q}_2(\alpha_2(\mathbf{X},t))\,\mathbf{Q}_1(\alpha_1(\mathbf{X},t))$ for different initial conditions converge with time to the inertial manifold $\mathcal{M}(t)$ as shown in Figure~\ref{fig:Qconverge}b. This is a direct consequence of convergence of the Protean basis vectors $\hat{\mathbf{e}}^\prime_i$ to the covariant Lyapunov vectors $\mathbf{a}_i$, and as shown Figure~\ref{fig:Qconverge}c, the standard deviation $\sigma_{Q_{ij}}(t)$ of the components $Q_{ij}(t)$ between different initial conditions exponentially converges as $\exp(-t/\tau_{\Delta\lambda})$ toward the precision of the numerical ODE solver. After an convergence time $t_c\approx 5\tau_{\Delta\lambda}$ the Protean basis vectors $\hat{\mathbf{e}}^\prime_i$ converge to the covariant Lyapunov vectors $\mathbf{a}_i$ to within $\sim 1\%$ error, and so Lagrangian velocity gradient statistics are gathered for $t-t_0>t_c$. We note that more sophisticated schemes to accelerate convergence of $\hat{\mathbf{e}}^\prime_i$ to $\mathbf{a}_i$ (and likewise $\alpha_i(t)$ to $\mathcal{M}(t)$) via back-substitution~\citep{Ginelli:2007aa,Froyland:2013aa,Noethen:2019aa} however these are not considered in this study; instead we rely upon then inherent convergence properties of (\ref{eqn:rotns}).

\subsection{Deformation in the Protean Frame}
\label{subsec:deform_Protean}

From (\ref{eqn:deform}), the principal components of the Protean deformation tensor $\mathbf{F}^\prime$ evolve exponentially in time as
\begin{equation}
F^\prime_{ii}(\mathbf{X},t)=\exp\left[\int_0^t\epsilon^\prime_{ii}(\mathbf{X},t^\prime)dt^\prime\right],\label{eqn:stretch}
\end{equation}
and so the ensemble average of the diagonal components of the deformation tensor grow as $\langle F^\prime_{ii}(t)\rangle \sim \exp(\langle\epsilon^\prime_{ii}\rangle t)$, and from (\ref{eqn:LyapunovQR}) the Lyapunov exponents of the flow are given by the temporal averages $\lambda_{\infty,i}=\langle\epsilon^\prime_{ii}\rangle$. Note that due to ergodicity, statistics of $\boldsymbol\epsilon^\prime$ may be gathered over multiple trajectories $\boldsymbol\epsilon^\prime(\mathbf{X},t)$  for $t-t_0>t_c$. As such, we refer to the diagonal components $\epsilon^\prime_{ii}(\mathbf{X},t)$ for $t>t_c$ as \emph{Lyapunov increments}. For volume-preserving flows $\sum_i\epsilon^\prime_{ii}(t)=0$, hence $\prod_i F^\prime_{ii}=\det{\mathbf{F}^\prime}=1$, whereas for compressible flows the constraints only apply in an averaged sense $\sum_i\langle\epsilon^\prime_{ii}\rangle=0$, $\prod_i\langle F^\prime_{ii}\rangle=1$. As shown in Appendix~\ref{app:strip}, the $\hat{\mathbf{e}}^\prime_1$ basis vector in the Protean frame converges to that of an infinitesimal 1D material strip, and $F'_{11}(\mathbf{X},t)$ quantifies the relative length of the strip at time $t$. 

There exist three distinct scenarios for the Lyapunov spectra $\lambda_{\infty,i}$ depending upon the number of positive or negative Lyapunov exponents; we assume the commonly-observed case~\citep{Girimaji:1990aa} for turbulent flows of two positive and one negative Lyapunov exponents and note that the extension to the other two cases is straightforward. 
From (\ref{eqn:deform}), the upper off-diagonal components of the deformation tensor evolve as
\begin{align}
F^\prime_{12}(t) &=F^\prime_{11}(t)\int_0^t dt^\prime\frac{\epsilon^\prime_{12}(t^\prime) F^\prime_{22}(t^\prime)}{F^\prime_{11}(t^\prime)},\label{eqn:F12}\\
F^\prime_{23}(t) &=F^\prime_{22}(t)\int_0^t dt^\prime\frac{\epsilon^\prime_{23}(t^\prime) F^\prime_{33}(t^\prime)}{F^\prime_{22}(t^\prime)},\label{eqn:F23}\\
F^\prime_{13}(t) &= F^\prime_{11}(t)\int_0^t dt^\prime\frac{\epsilon^\prime_{12}(t^\prime)F^\prime_{23}(t^\prime)+\epsilon^\prime_{13}(t^\prime)F^\prime_{33}(t^\prime)}{F^\prime_{11}(t^\prime)},\label{eqn:F13}
\end{align}
and the lower off-diagonal terms are all zero; $F'_{ij}=0$ for $i>j$. Hence the diagonal components of $\mathbf{F}'(\mathbf{X},t)$ all grow exponentially in time, whereas the off-diagonal components that objectively encode shear and vorticity (quantified by $\epsilon^\prime_{ij}$ with $j>i$) initially grow algebraically as $F_{ij}^\prime(t)\sim t^r$ at times shorter than then Lyapunov time $t<\tau_{\lambda}\equiv 1/\lambda_{\infty,1}$ and as the product $F_{ij}^\prime(t)\sim t^r \exp(\lambda_{\infty,1}t)$ for $t>\tau_\lambda$, as described by \cite{Lester:2018aa}.

\subsection{Stochastic Model}
\label{subsec:stochastic}

To develop a stochastic model for the evolution of the components of the Protean deformation tensor $\mathbf{F}^\prime$, we consider how the components of the velocity gradient tensor $\boldsymbol\epsilon^\prime$ evolve in time. In \S\ref{sec:decorrelation} and \citep{Dentz:2025aa} we establish that although unsteady flows such as HIT that are non-separable in space and time exhibit Lagrangian velocity statistics that are non-Markovian and non-Fickian, these are rendered Fickian on timescales longer than $\tau_c$; hence evolution of $\boldsymbol\epsilon$ be captured via a simple Brownian model. Several studies~\citep{Girimaji:1990aa} show that for, e.g., HIT, the $\epsilon_{ij}$ have finite mean $\langle\epsilon_{ij}\rangle$ and variance $\sigma_{\epsilon_{ij}}^2$ and the autocorrelation function of the components of $\boldsymbol\epsilon$ decays approximately exponentially in time at a rate given by the Kolmogorov time scale $\tau_\eta\equiv\sqrt{\nu/\varepsilon}$, where $\nu$ is the kinematic viscosity and $\varepsilon$ the turbulent dissipation rate~\citep{Yu:2010aa,Meneveau:2011aa}.

Although the velocity gradient components in, e.g., turbulent flows typically follow non-Gaussian statistics~\citep{Meneveau:2011aa}, the PDFs of $\epsilon_{ij}$ exhibit bounded variance. Furthermore, the log-stretch $\xi_{ii}\equiv \ln F_{ii}^\prime$
in (\ref{eqn:stretch}) follow an additive process with respect to $\epsilon^\prime_{ii}$.
Assuming that the Protean velocity gradient components $\epsilon_{ij}^\prime$ also exhibit bounded variance, the log-stretches $\xi_{ii}$ are asymptotically Gaussian distributed due to the central limit theorem. As a consequence, the specific distributions of the $\epsilon_{ij}'$ are not important. We assume they are Gaussian distributed and correlated on a characteristic correlation time scale. Furthermore, we assume that the time series of the $\epsilon_{ij}'(t)$ are stationary because they converge asymptotically to their mean values. Thus, we assume that the $\epsilon_{ij}'(t)$ are Gaussian and stationary random processes. The Doob theorem \citep{doob1942brownian} then states that the $\epsilon_{ij}^\prime(t)$ evolve via the multivariate Ornstein-Uhlenbeck (OU) process\citep{Gardiner:2004aa}
\begin{equation}
    d\hat{\boldsymbol\epsilon}=-\mathbf{P}(\hat{\boldsymbol\epsilon}-\boldsymbol\lambda)dt+\mathbf{V}\,d\mathbf{W}_t,\label{eqn:OU}
\end{equation}
where $\hat{\boldsymbol\epsilon}$ is the $d(d-1)$-vector of non-zero components $\epsilon^\prime_{ij}$ ($i\leqslant j$),
$\boldsymbol\lambda\equiv\langle\boldsymbol\epsilon\rangle$, $\mathbf{P}$ and $\mathbf{V}$ are $d(d-1)\times d(d-1)$ symmetric constant matrices that respectively characterise drift rate and diffusion, and $d\mathbf{W}_t$ is a $d(d-1)-$dimensional Wiener process. As such, the components $\hat{\epsilon}_{i}$ of $\hat{\boldsymbol\epsilon}$ decorrelate exponentially in time according to the correlation matrix
\begin{equation}
    \mathbf{R}_{\hat{\epsilon}}(\tau)=\exp(\mathbf{P\,\tau)}\mathbf{R}_{\hat\epsilon}(0),\label{eqn:OUcorr}
\end{equation}
where
\begin{equation}
    R_{\hat\epsilon,ij}(\tau)\equiv\frac{\langle(\hat{\epsilon}_{i}(t)-\lambda_i)(\hat\epsilon_{j}(t+\tau)-\lambda_j)\rangle}{\sigma_{\hat\epsilon_{i}}\sigma_{\hat\epsilon_{j}}}
    ,
\end{equation}
where $i,j=1:d(d-1)$ and $\sigma^2_{\hat{\epsilon_i}}$ is the variance of $\hat{\epsilon}_i$. From (\ref{eqn:OUcorr}), the drift matrix $\mathbf{P}$ can be obtained from the correlation matrix $\mathbf{R}_{\hat\epsilon}(\tau)$ as
\begin{equation}
    \mathbf{P}=\frac{1}{\tau}\ln\left(\mathbf{R}_{\hat\epsilon}(\tau)\mathbf{R}_{\hat\epsilon}(0)^{-1}\right),
\end{equation}
where $\mathbf{R}_{\hat\epsilon}(0)$ is the correlation matrix and the dispersion matrix $\mathbf{V}$ can be determined from the covariance matrix $\boldsymbol\Sigma=\mathbf{V}\mathbf{V}^\top$ via Cholesky decomposition, where $\boldsymbol\Sigma$ is given by
\begin{equation}
    \boldsymbol\Sigma=\left(-\mathbf{P}\mathbf{R}_{\hat\epsilon}(0)^\top+\mathbf{R}_{\hat\epsilon}(0)\mathbf{P}^\top\right).
\end{equation}
Under the further assumption that only the diagonal components $\epsilon_{ii}^\prime$ are correlated with each other, the log-stretches $\xi_{ii}\equiv F_{ii}^\prime(t)=\int_0^t \epsilon^\prime_{ii}(s)ds$ evolve with mean $\langle\xi_{ii}\rangle=\lambda_{\infty,i}t$, and the covariance matrix $\mathbf{R}_\xi(t)$ for $\xi_{ii}$ is given as
\begin{equation}
    \mathbf{R}_\xi(t)=\int_0^t\int_0^t \mathbf{R}_{ii}(\tau_1-\tau_2)d\tau_1d\tau_2=2\int_0^t (t-\tau)\exp\left(\mathbf{P}_{ii}\tau\right)\mathbf{R}_{ii}(0)d\tau,
\end{equation}
where $\mathbf{R}_{ii}(\tau)$, $\mathbf{P}_{ii}$ are the symmetric $d\times d$ submatrices of $\mathbf{R}_{\hat\epsilon}(\tau)$, $\mathbf{P}$ that correspond to the diagonal $\epsilon^\prime_{ii}$ components. For $t\gg\tau_c$, then from the central limit theorem
\begin{equation}\label{eq:Dxi}
    \mathbf{R}_\xi(t)\approx 2t\int_0^\infty \mathbf{R}_{ii}(\tau)d\tau\equiv 2\mathbf{D}_\xi\, t,
\end{equation}
where $\mathbf{D}_\xi=-\mathbf{P}^{-1}\mathbf{R}_{ii}(0)$ is the effective diffusivity tensor. Hence, although $\epsilon^\prime$ is correlated at short times, $\xi_{ii}(t)$ evolve according to Brownian motion at long times with renormalized diffusivity $\mathbf{D}_\xi$ as
\begin{equation}
d\xi_{ii}=\lambda_{\infty,i}\,dt+\sqrt{2}B_{ij} \,  dW_j(t),\label{eqn:brownian}
\end{equation}
where $\mathbf{B}\mathbf{B}^\top=\mathbf{D}_{\xi}$ and $dW_j(t)$ with $j=1:d$ are independent Brownian motions.
Hence the log-stretches $\xi_{ii}(t)$ are Gaussian-distributed with mean $\lambda_{\infty,i} t$, variance $\sigma_{ii}^2 t$ and the vector of log-stretches $\boldsymbol\xi=(\xi_{11},\xi_{22},\xi_{33})$ has mean $\boldsymbol\lambda_\infty\,t=(\lambda_{\infty,1},\lambda_{\infty,2},\lambda_{\infty,3})t$ and correlation matrix $\boldsymbol\Sigma\,t$, and the principal stretches $F^\prime_{ii}(t)$ are log-normally distributed with log-mean $\boldsymbol\lambda\,t$ and log-covariance matrix given by $\boldsymbol\Sigma\,t$. As the Lyapunov exponents satisfy $\lambda_{\infty,1}>\lambda_{\infty,2}>\lambda_{\infty,3}$, then the integrals in (\ref{eqn:F12})-(\ref{eqn:F23}) converge to constants at a rate controlled by the spectral gap time $\tau_{\Delta\lambda}$ as 
\begin{align}
A_{12}(\mathbf{X},t) & \equiv \frac{F_{12}(\mathbf{X},t)}{F_{11}(\mathbf{X},t)} \rightarrow a_{12}(\mathbf{X}),\label{eqn:A12}\\
A_{23}(\mathbf{X},t) & \equiv \frac{F_{23}(\mathbf{X},t)}{F_{22}(\mathbf{X},t)} \rightarrow a_{23}(\mathbf{X}),\label{eqn:A23}\\
A_{13}(\mathbf{X},t) & \equiv \frac{F_{13}(\mathbf{X},t)}{F_{11}(\mathbf{X},t)} \rightarrow a_{13}(\mathbf{X}).\label{eqn:A13}
\end{align}
%
 From (\ref{eqn:brownian})-(\ref{eqn:A13}), a closed form expression for the rescaled right Cauchy-Green tensor $\mathbf{C}^\star(\mathbf{X},t)\equiv\mathbf{C}^\prime(\mathbf{X},t)/F^\prime_{11}(\mathbf{X},t)^2$ is
\begin{equation}\mathbf{C}^\star(\mathbf{X},t)=
\left(\begin{array}{ccc}
1 & a_{12} & a_{13}\\
a_{12} & a_{12}^2+m_{22}^2 & a_{12}a_{13}+a_{23}m_{22}\\
a_{13} & a_{12}a_{13}+a_{23}m_{22} & a_{13}^2+a_{23}^2+m_{33}^2
\end{array}
\right),
\end{equation} 
where $m_{ii}(\mathbf{X},t)\equiv F_{ii}(\mathbf{X},t)/F_{11}(\mathbf{X},t)$ and $m_{ii}\rightarrow 0$ for $t\gg \tau_{\Delta\lambda}$ and $i=2,3$. As the Cauchy-Green tensor $\mathbf{C}(\mathbf{X},t)$ for $t\gg \tau_\lambda$ converges to a steady matrix multiplied by $F_{11}(\mathbf{X},t)^2$, local fluid deformation is dominated by stretching associated with $\epsilon^\prime_{11}$. Furthermore, the leading eigenvalue of $\mathbf{C}(\mathbf{X},t)$ rapidly converges to $\sigma_1(\mathbf{X},t)\approx F_{11}(\mathbf{X},t)^2(1+a_{12}(\mathbf{X})^2+a_{13}(\mathbf{X})^2+a_{23}(\mathbf{X})^2)$, and so the finite-time Lyapunov exponent (FTLE) $\lambda(\mathbf{X},t)$ also evolves as
\begin{equation}
\begin{split}
\lambda_1(\mathbf{X},t)&\equiv\frac{1}{2t}\ln\sigma_1(\mathbf{X},t)\approx\frac{1}{t}\xi_{11}(\mathbf{X},t)+\frac{1}{2t}\ln(1+a_{12}(\mathbf{X})^2+a_{13}(\mathbf{X})^2+a_{23}(\mathbf{X})^2)\\
&\rightarrow\lambda_{\infty,1}+\frac{\sigma_{\lambda}^2\zeta(t)}{\sqrt{t}},\label{eqn:FTLE}
\end{split}
\end{equation}
where $\zeta(t)$ is a white noise with zero mean and unit variance where $\langle\zeta(t)\zeta(t^\prime)\rangle=\exp(-|t-t^\prime|/\tau_c)$. Averaging over $N$ trajectories (denoted as $\langle\cdot\rangle_N$) yields faster convergence of the FTLE to the maximum Lyapunov exponent $\lambda_{\infty,1}$ as
\begin{equation}
\langle\lambda_1(\mathbf{X},t\rangle_N=\lambda_{\infty,1}+\frac{\sigma_{\lambda}^2}{\sqrt{N t}}\rightarrow\lambda_{\infty,1}.\label{eqn:FTLEmean}
\end{equation}
Under the assumption of finite variance of $\epsilon^\prime_{ij}$, and non-zero correlation only between the diagonal components $\epsilon^\prime_{ii}$, the simple Brownian model (\ref{eqn:brownian}) for evolution of the log-stretches $\xi_{ii}\equiv \mathbf{F}_{ii}^\prime$ provide a tractable stochastic model for the evolution of the Cauchy-Green tensor, FTLEs and the Lyapunov spectra from Lagrangian velocity gradient data. In the following section we apply this method to numerical data for 2D and 3D unsteady random flows.

\section{Numerical Examples}
\label{sec:numerics}

\subsection{2D Kraichnan Flow}
\label{subsec:Kraichnan}

\begin{figure}
\centering
\begin{tabular}{c c c}
\includegraphics[height=0.3\columnwidth]{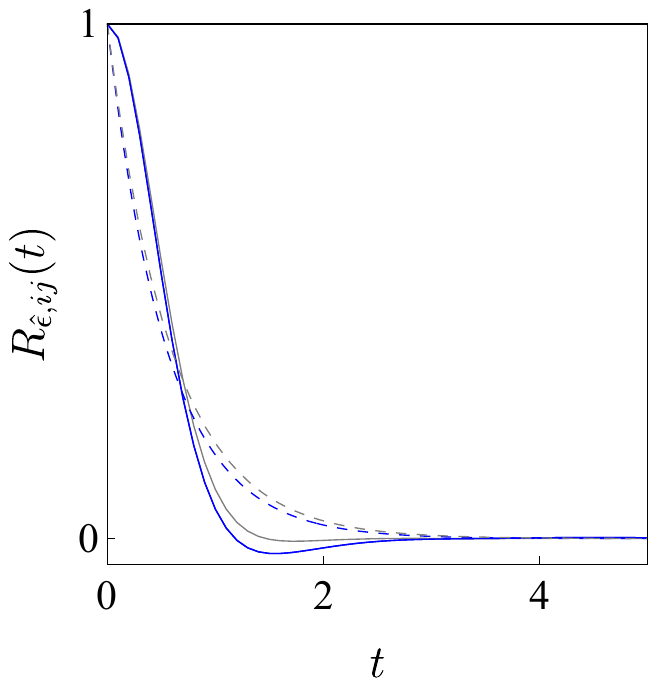}&
\includegraphics[height=0.3\columnwidth]{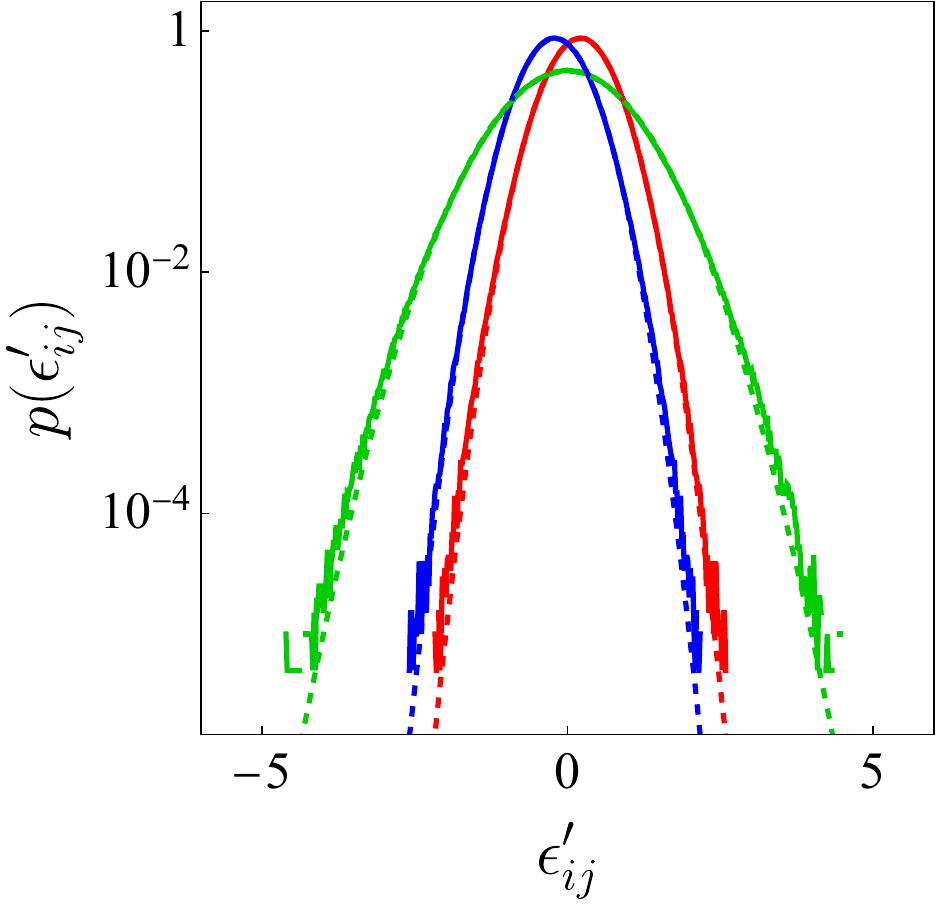}&
\includegraphics[height=0.3\columnwidth]{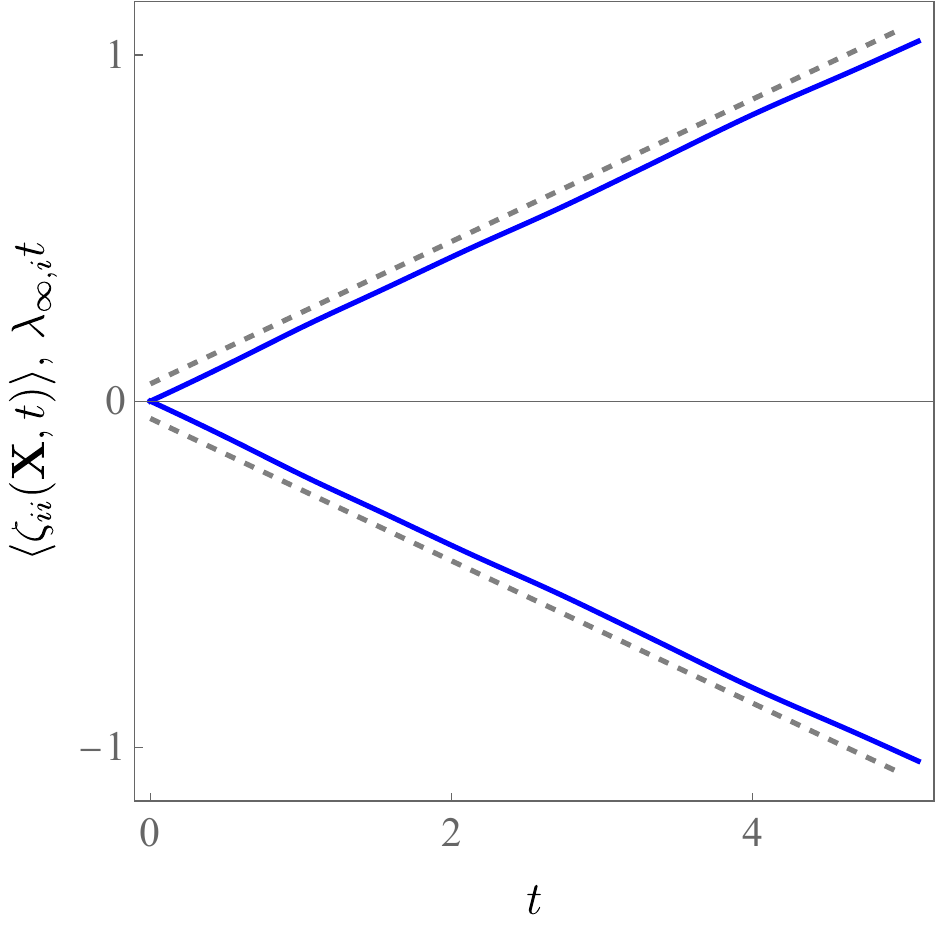}\\
(a) & (b) & (c)\\
\includegraphics[height=0.3\columnwidth]{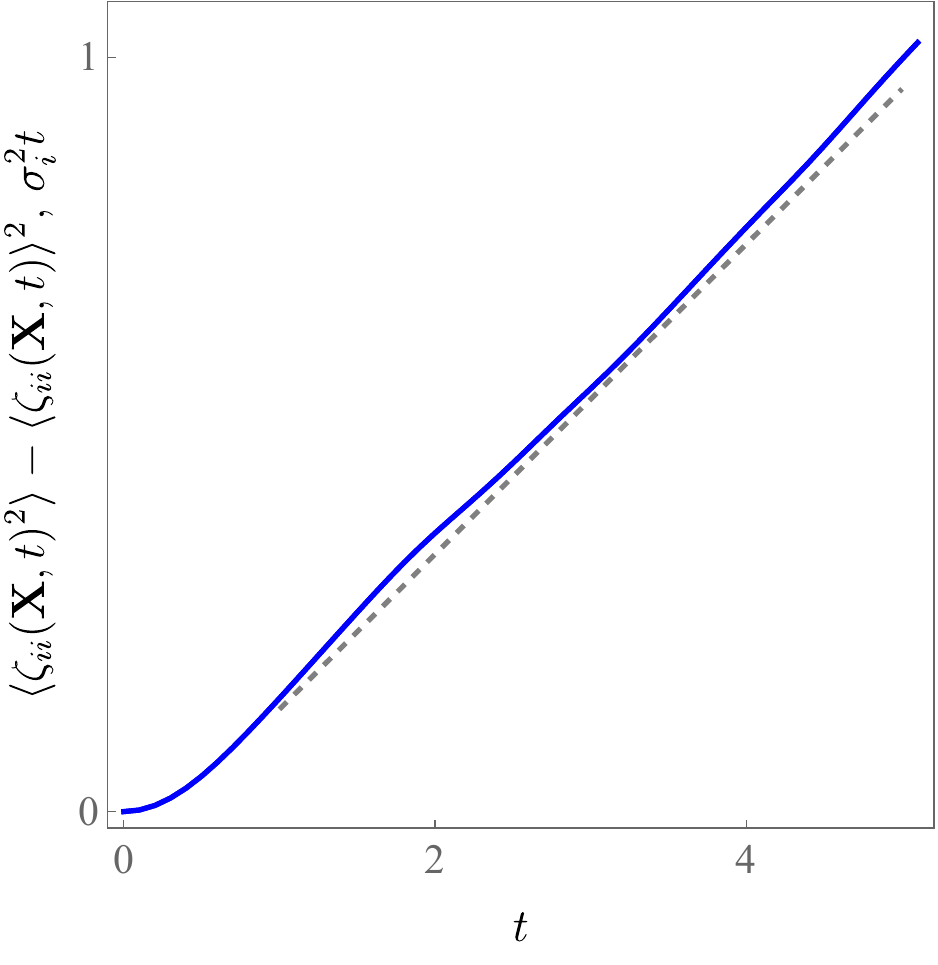}&
\includegraphics[height=0.3\columnwidth]{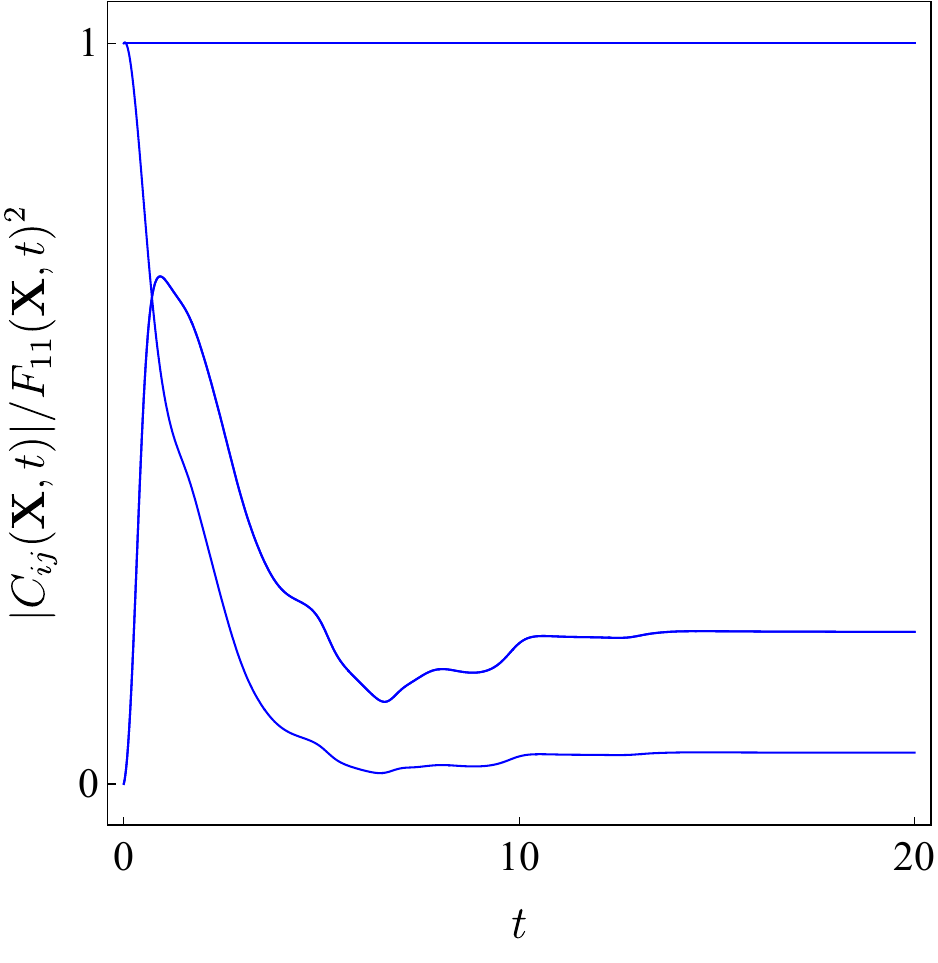}&
\includegraphics[height=0.3\columnwidth]{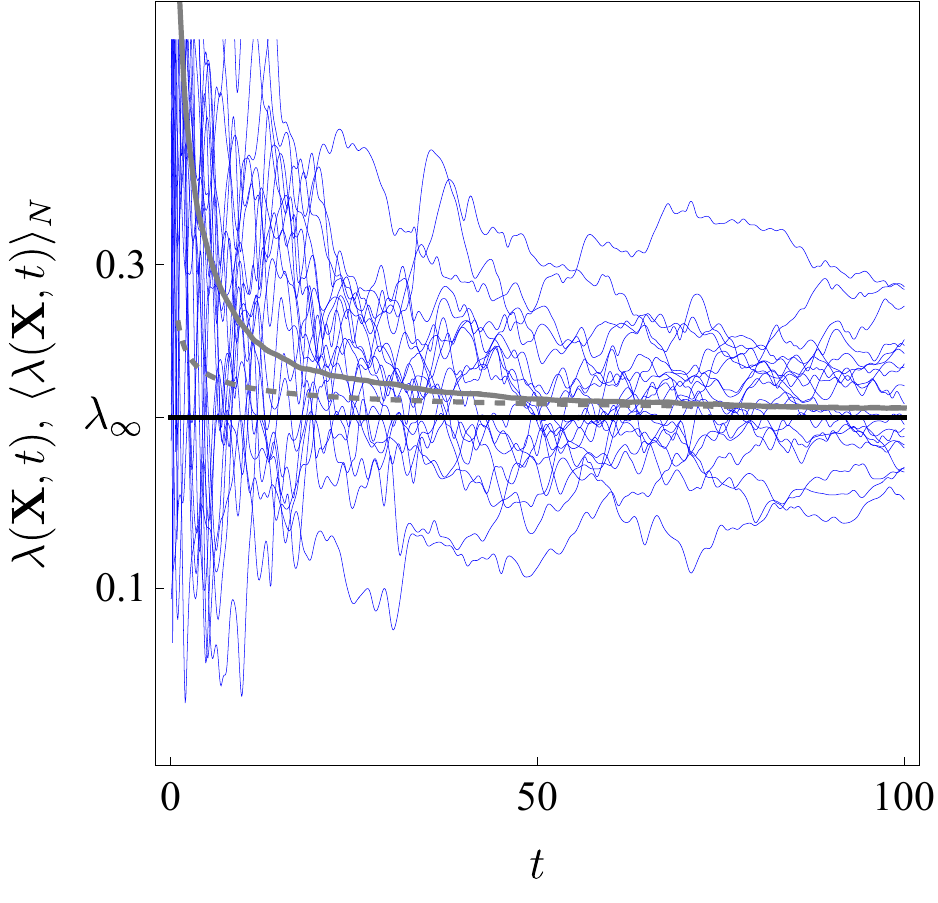}\\
(c) & (d) & (e)
\end{tabular}
\caption{
(a) Autocorrelation functions (solid lines) and exponential fits (dashed lines) for diagonal ($\epsilon_{11}^\prime$, $\epsilon_{22}^\prime$, blue) and non-zero off-diagonal ($\epsilon_{12}^\prime$, grey) components of the Lagrangian velocity gradient tensor $\boldsymbol\epsilon$ of the 2D Kraichnan flow. (b) PDFs of the non-zero components (red $\epsilon^\prime_{11}$; blue $\epsilon^\prime_{22}$; green $\epsilon^\prime_{12}$) of the Protean velocity gradient tensor. Dashed lines represent fits of the Gaussian distribution to these PDFs. Evolution of (c) ensemble mean $\langle\zeta_{ii}(t)\rangle$ and (d) ensemble variance $\langle\zeta_{ii}(t)^2\rangle-\langle\zeta_{ii}(t)\rangle^2$ of log-stretches (solid blue lines) over $10^3$ trajectories and respective analytic solutions $\lambda_{\infty,i}t$, $\sigma_{ii}^2 t$ (dashed gray lines, offset fo clarity) for $i=1,2$. (e) Convergence of scaled components $|C_{ij}(\mathbf{X},t)|/F_{11}(\mathbf{X},t)^2$ of the Cauchy-Green tensor to a constant value for a typical particle trajectory on timescale $t\gg\tau_{\Delta\lambda}$. (f) Evolution of FTLE $\lambda(\mathbf{X},t)$ (light blue lines) over 30 sample trajectories and ensemble averaged FTLE $\langle\lambda(\mathbf{X},t)\rangle_N$ with Lagrangian time $t$ (black dashed line) toward Lyapunov exponent $\lambda_\infty$ (solid black line). The analytic expression (\ref{eqn:FTLEmean}) for $\langle\lambda(\mathbf{X},t)\rangle_N$ (solid gray line) is different to the numerical solution at short times due to the finite-time convergence shown in (e).}\label{fig:Kraichnan}
\end{figure}

To illustrate and test the stochastic model developed in \S\S~\ref{subsec:stochastic}, we first consider fluid deformation in the 2D Kraichnan model flow described in Appendix~\ref{app:Kraichnan} with finite correlation time $\tau_c=1$. Here, the Lagrangian velocity gradient tensor $\boldsymbol\epsilon(\mathbf{X},t)$ is computed along $10^3$ distinct particle trajectories with random $\mathbf{X}$ for an advection time of $10^4$ (thus generating $10^7$ independent observations) and the Protean orientation angle $\alpha_3(\mathbf{X},t)$ given by (\ref{eqn:alpha3}) with initial condition $\alpha_{3,0}=0$ is solved numerically to accuracy $10^{-10}$ using a variable time step Runge-Kutta scheme. Statistics of $\boldsymbol\epsilon^\prime$ are gathered after the convergence time $4\tau_{\Delta\lambda}\approx 10$, and the results are summarised in Figure~\ref{fig:Kraichnan}. Due to the divergence free condition $\sum_i\epsilon^\prime_{ii}=0$, the diagonal components $\epsilon^\prime_{11}$, $\epsilon^\prime_{22}$ are found to be perfectly anti-correlated $(\rho(\epsilon^\prime_{11},\epsilon^\prime_{22})=-1)$, and effectively independent of the off-diagonal term $\rho(\epsilon^\prime_{11},\epsilon^\prime_{12})\approx 0.00076$. As shown in Figure~\ref{fig:Kraichnan}a, these components decorrelate near-exponentially in time, and the diagonal components have an almost identical temporal correlation structure. Computation of the effective diffusion tensor $\mathbf{D}_\xi$, see Eq. \eqref{eq:Dxi}, confirms that the diagonal and off-diagonal components of $\boldsymbol\epsilon^\prime$ are effectively independent (with $|D_\xi(\epsilon^\prime_{11},\epsilon^\prime_{12})|<0.01$), and the diagonal components are anti-correlated $D_\xi(\epsilon^\prime_{11},\epsilon^\prime_{11})=-D_\xi(\epsilon^\prime_{11},\epsilon^\prime_{22})= \tau_{c,11}=\tau_{c,22}\approx 0.527$, and $D_\xi(\epsilon^\prime_{12},\epsilon^\prime_{12})\approx 0.610$. As expected, the Lyapunov spectrum $(\lambda_{\infty,1},\lambda_{\infty,2},\lambda_{\infty3})$ is perfectly symmetric with $\lambda_{\infty,1}=-\lambda_{\infty,3}\approx 0.2056$ (hence $\tau_\Delta\lambda\approx 2.431$), and the shear term has nearly zero mean ($\lambda_{\infty,2}\approx -0.00011$), and as shown in Figure~\ref{fig:Kraichnan}b, the components $\epsilon^\prime_{ij}$ are all Gaussian distributed.

Hence, the deformation structure of the 2D Kraichnan flow is remarkably simple. The Lyapunov increments $\epsilon_{11}^\prime$, $\epsilon_{22}^\prime$ are perfectly anti-correlated and essentially independent of the shear component $\epsilon_{12}^\prime$, and all of these components are Gaussian distributed with near-exponential temporal decorrelation, meaning that the OU process (\ref{eqn:OU}) provides an accurate representation of the Lagrangian deformation dynamics, which also decouples between and $\epsilon_{11}^\prime$ and $\epsilon_{12}^\prime$. As shown in Figure~\ref{fig:Kraichnan}c,d, this also leads to a remarkably simple deformation dynamics, where the ensemble statistics of the log-stretches $\zeta_{ii}$ quickly converge to their theoretical estimates from the OU process. Figure~\ref{fig:Kraichnan}e shows that for a typical trajectory, the scaled components of the Cauchy-Green tensor all converge to constant values on the timescale $t\gg\tau_{\Delta\lambda}$, hence at long times the deformation is controlled solely by line stretching as quantified by $F_{11}(\mathbf{X},t)$ Similarly, Figure~\ref{fig:Kraichnan}f shows that the ensemble averaged FTLEs rapidly converge to $\lambda_\infty$, in accordance with (\ref{eqn:FTLEmean}) for $t\gg\tau_{\Delta\lambda}$.

\subsection{Homogeneous Isotropic Turbulence}
\label{subsec:HIT}

For further illustration of the stochastic model developed in \S\S~\ref{subsec:stochastic} and to eludicate the structure of turbulent deformation, we utilise direct numerical simulation (DNS) data hosted as part of the Johns Hopkins Turbulence Database (JHTU) of forced isotropic turbulence at Taylor-scale Reynolds number $Re_\lambda\approx433$ computed using 1024$^3$ nodes via a pseudo-spectral method. Tracer particle position, velocity and velocity gradient data is recorded over randomly sampled points at fixed spatial locations and fixed times, as well as for 10$^3$ fluid tracer trajectories with random initial locations for a time period of 10 time units (corresponding to around five eddy turnover times $T_L$). This data allows for characterisation of both the Eulerian and Lagrangian spatio-temporal correlation structures, as well as the Eulerian velocity statistics and the evolution of the Lagrangian velocity gradient tensor.

\begin{figure}
\centering
\begin{tabular}{c c c}
\includegraphics[height=0.3\columnwidth]{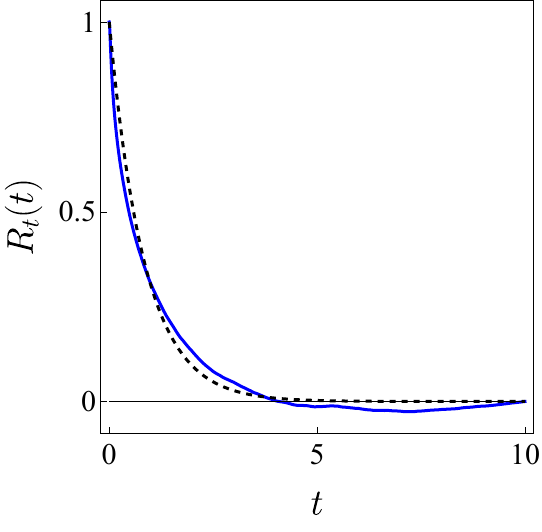}&
\includegraphics[height=0.3\columnwidth]{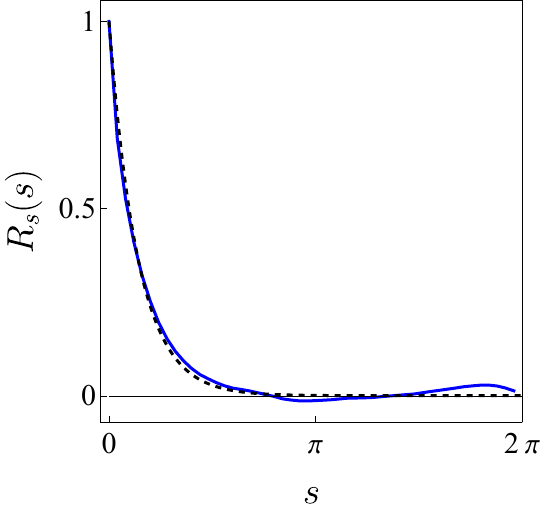}&
\includegraphics[height=0.3\columnwidth]{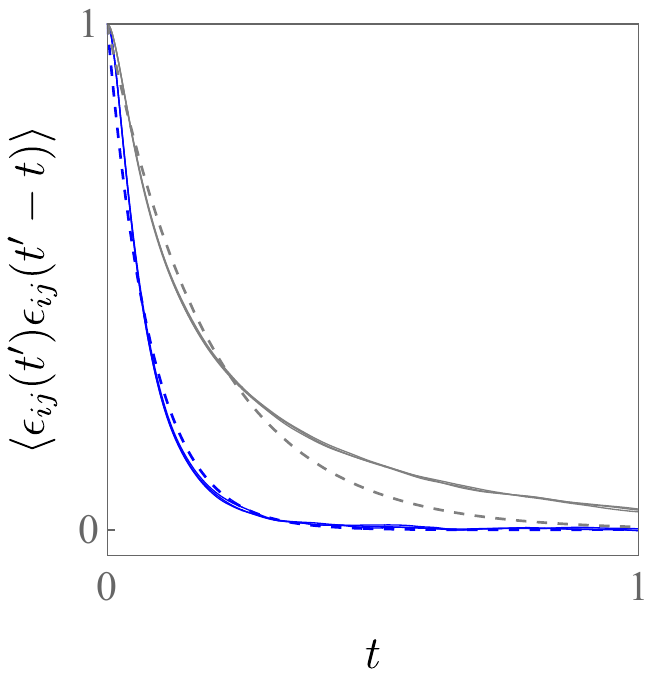}\\
(a) & (b) & (c) \\
\includegraphics[height=0.3\columnwidth]{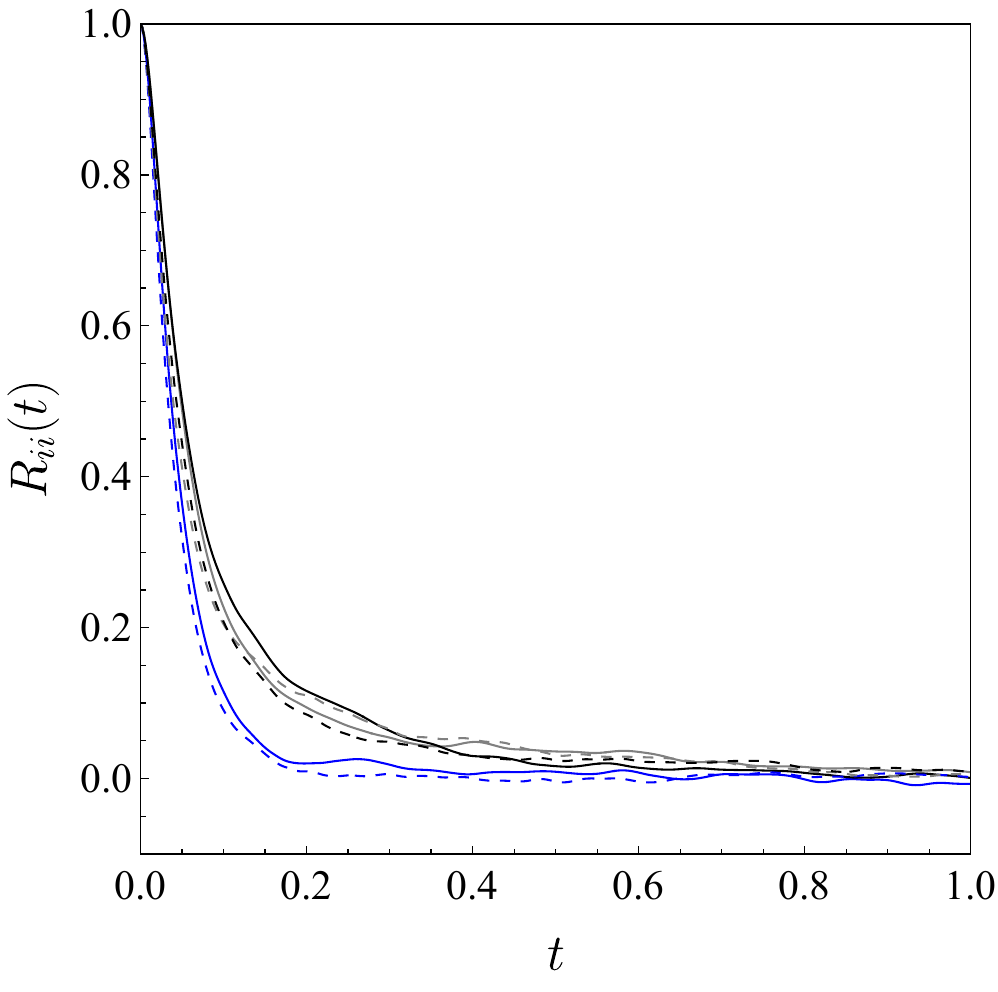}&
\includegraphics[height=0.3\columnwidth]{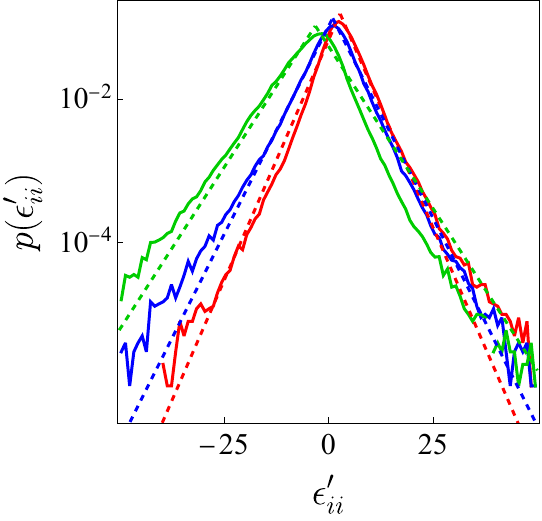}&
\includegraphics[height=0.3\columnwidth]{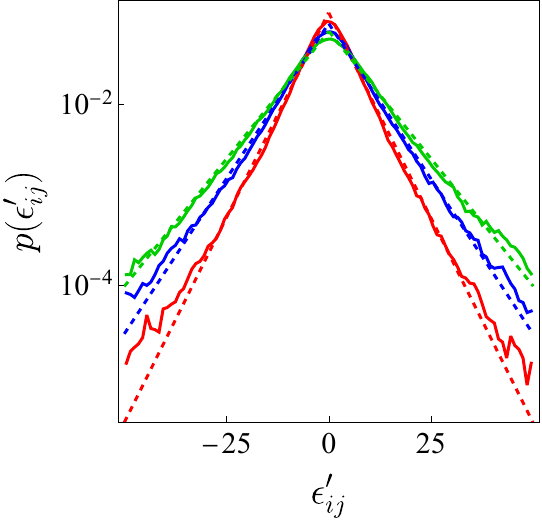}\\
(d) & (e) & (f)
\end{tabular}
\caption{
(a) Temporal and (b) spatial autocorrelation functions for the Lagrangian velocity magnitude. Dashed lines represent fitted respective temporal and spatial autocorrelation functions $R_t(t)=\exp(-t/t_d)$, $R_s(s)=\exp(-s.s_d)$, where $t_d=0.844$ [s], $s_d=0.439$ [m]. 
(c) Autocorrelation functions (solid lines) and exponential fits (dashed lines) for diagonal (blue) and off-diagonal (grey) components of the Lagrangian velocity gradient tensor $\boldsymbol\epsilon$. (d) Autocorrelation functions (solid lines) and exponential fits (dashed lines) for all non-zero components of the Protean Lagrangian velocity gradient tensor $\boldsymbol\epsilon^\prime$. PDFs of the (e) diagonal $\epsilon^\prime_{ii}$ and (f) off-diagonal $\epsilon^\prime_{ii}$ components of the Protean velocity gradient tensor. Dashed lines represent fits of the Laplace distribution to these PDFs.}\label{fig:velpdfs}
\end{figure}

Spatial and temporal decorrelation of the velocity magnitude $v$ is shown in Figure~\ref{fig:velpdfs}(a), (b), indicating decorrelation is approximately exponential (first order and linear) in both space and time. The transition to negative correlation with finite time or distance shown in Figure~\ref{fig:velpdfs}(a), (b) is due to sweeping behaviour of the local flow. The fitted decorrelation temporal and spatial scales are respectively $\tau_c=0.844$, $\ell_v=0.439$, which in conjunction with the mean velocity $\langle v\rangle=1.091$, yields the global Kubo number
\begin{equation}
\kappa\equiv \langle v\rangle\frac{\tau_c}{\ell_v}\approx 2.0975,
\end{equation}
indicates that spatial velocity decorrelation occurs approximately twice as fast as temporal decorrelation. The Eulerian velocity PDF $p_e(v)$ (not shown) is well-fitted by the Nakagami distribution
\begin{equation}
p_e(v)=\frac{2}{\Gamma(\beta)}\left(\frac{\beta}{\alpha}\right)^\beta v^{2\beta-1}\exp\left(-\frac{v^2\beta}{\alpha}\right),\label{eqn:Nakagami}
\end{equation}
where $\beta=1.4903$, $\alpha=1.3854$ are statistical parameters of the distribution, which recover the mean Eulerian velocity as $\langle v\rangle=\sqrt{\alpha/\beta}\Gamma(\beta+1/2)/\Gamma(\beta)\approx1.0902$. For small velocities, this distribution has the power-law scaling $p_e(v)\sim v^{2\beta-1}=v^{1.981}$. Although this velocity scaling ($1<\beta<2$) is typically associated with non-Fickian transport in e.g. steady flows, the finite Kubo number ($\kappa\approx 2$) means that temporal decorrelation plays a significant role in resetting the velocity along pathlines, leading a transition to Fickian behaviour for times greater than $\tau_c$.

The autocorrelation structure of the Lagrangian velocity gradient $\boldsymbol\epsilon$ along pathlines is shown in Figure~\ref{fig:velpdfs}a. The diagonal components $\epsilon_{ii}$ decorrelate with correlation time $\tau_{\epsilon_{ii}}=0.0802$, which is very similar to the Kolmogorov time $\tau_\eta=0.0844$ reported for this flow~\citep{Li:2008aa}. Conversely, the off-diagonal components $\epsilon_{ij}$, $j>i$ decorrelate more slowly, with correlation time $\tau_{\epsilon_{ij}}=0.211-0.216$, typical for persistent vortices in HIT~\citep{Nagata:2020aa}. The autocorrelation structure of the Protean Lagrangian velocity gradient $\boldsymbol\epsilon^\prime$ shown in Figure~\ref{fig:velpdfs}b is quite different, where the $\epsilon_{ij}$, 

Numerical solution of the ODE (\ref{eqn:rotns}) and application of the rotation $\mathbf{Q}(\mathbf{X},t)$ to the velocity gradient data along pathlines generates the Protean velocity gradient tensor $\boldsymbol\epsilon^\prime(\mathbf{X},t)$. The Lagrangian correlation structure of the diagonal velocity gradient components $\epsilon^\prime_{ii}$ is shown in Figure~\ref{fig:velpdfs}d, which in accordance with \eqref{eqn:OUcorr}, indicates that these components decorrelate approximately exponentially in Lagrangian time. Interestingly, the individual elements of $\boldsymbol\epsilon^\prime$ have different correlation times $\tau_{c,ij}$ that are roughly paired as $\tau_{c,22}\approx\tau_{c,12}\approx 0.046$, $\tau_{c,11}\approx\tau_{c,13}\approx 0.072$, $\tau_{cc,23}\approx\tau_{c,33}\approx 0.062$. Although these correlation times have not yet fully converged (in terms of dependence upon the number of particle trajectories), they appear to exhibit these persistent pairings and differences, which at the stage are not fully understood.

The Protean velocity gradient PDFs shown in Figure~\ref{fig:velpdfs}(e,f) (sampled at fixed temporal increments $\tau_c$ along a pathline) and as shown, all components are well-fitted by a Laplace distribution. From the ensemble averages of the diagonal components $\epsilon^\prime_{ii}$, the Lyapunov exponents are $(\lambda_{\infty,1},\lambda_{\infty,2},\lambda_{\infty,3})=(\langle\epsilon^\prime_{11}\rangle,\langle\epsilon^\prime_{22}\rangle,\langle\epsilon^\prime_{33}\rangle,)=(3.047,0.904,-3.961)$, consistent with previous studies~\citep{Girimaji:1990aa} that observe $\lambda_{\infty,2}$ to be positive and around one third the magnitude of $\lambda_{\infty,1}$ in of turbulent flows. These Lyapunov exponents correspond to a minimum spectral gap of $\lambda_{\infty,1}-\lambda_{\infty,2}=2.143$, hence the Lagrangian advection time $T=10$ is significantly longer than the convergence time $t_\lambda=1/(\lambda_{\infty,1}-\lambda_{\infty,2})=0.466$, indicating that Protean frame basis vectors $\mathbf{q}_i$ have converged to the Lyapunov vectors $\mathbf{a}_i$. As expected, the off-diagonal components $\epsilon^\prime_{ij}$ all have negligible ensemble averages ($|\langle\epsilon^\prime_{ij}\rangle|<10^{-4}$).

All of the non-zero velocity gradient components have finite variance, with the diagonal components having lower variance ($\sigma_{11}^2$=20.36, $\sigma_{22}^2$=27.42, $\sigma_{33}^2$=43.49) than their off-diagonal counterparts ($\sigma_{12}^2$=45.90, $\sigma_{13}^2$=80.88, $\sigma_{23}^2$=120.08). The correlation matrix for the six non-zero velocity gradient components $\boldsymbol\Sigma$ shows that the all of the velocity gradient components are independent (with cross correlation $|\rho|<10^{-3}$)) except for the diagonal velocity gradient components $\epsilon_{ii}^\prime$ which are found to be weakly negatively correlated due the incompressibility constraint, where the correlations between components are $\rho_{11,22}$=-0.091, $\rho_{11,33}$=-0.612, $\rho_{22,33}=-0.732$. 

\begin{figure}
\centering
\begin{tabular}{c c}
\includegraphics[width=0.4\columnwidth]{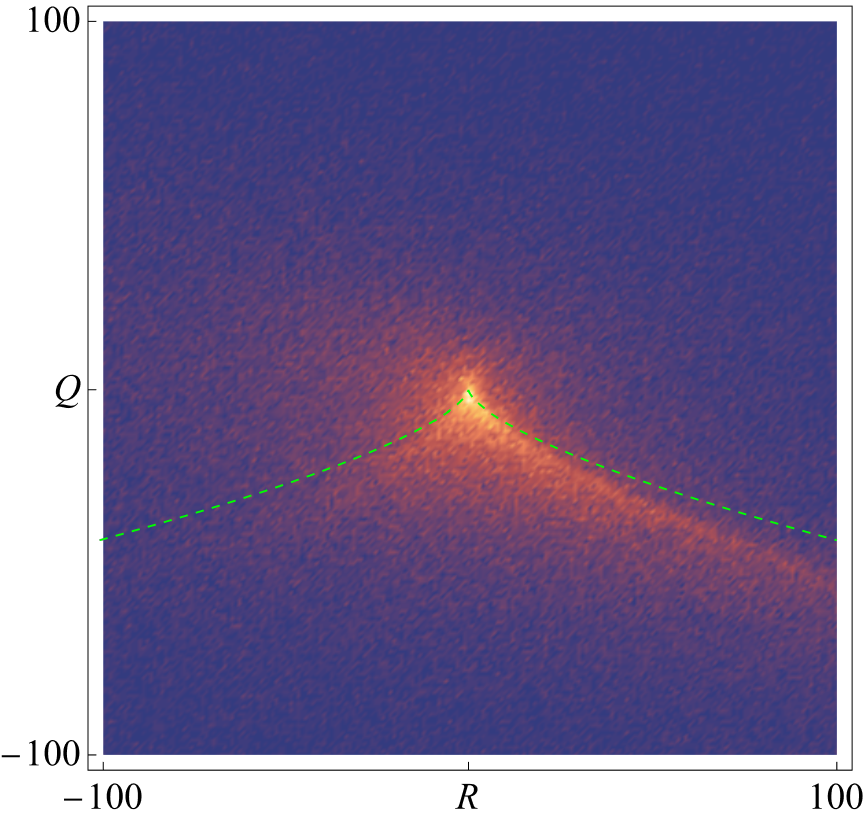}&
\includegraphics[width=0.4\columnwidth]{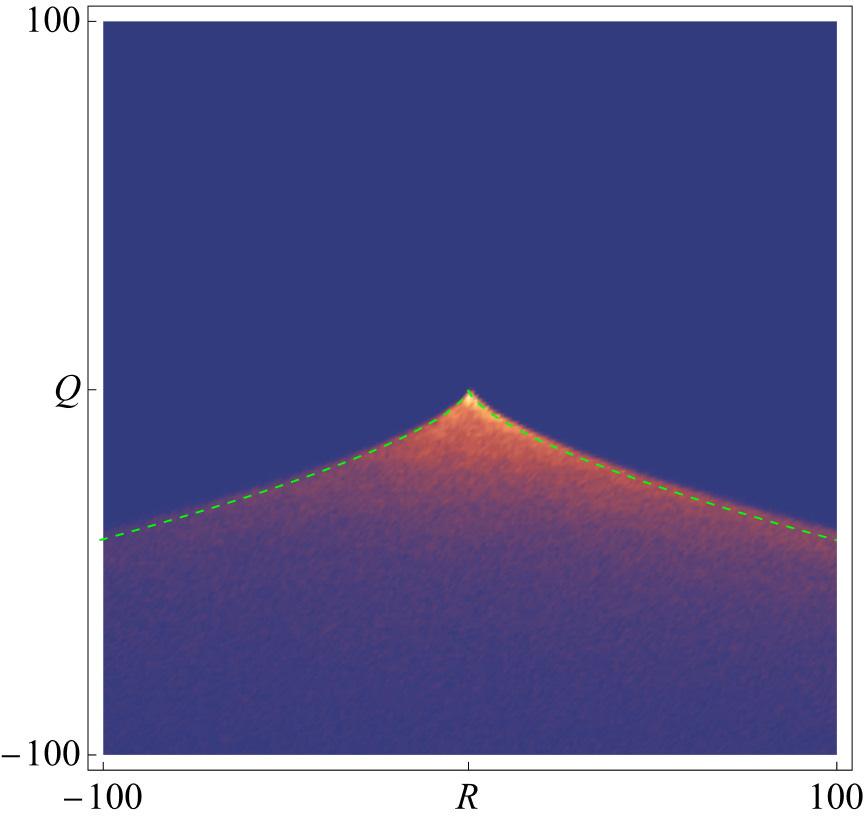}\\
(a) & (b)
\end{tabular}
\caption{Joint probability distribution of the $Q$, $R$ invariants of the velocity gradient tensor $\boldsymbol\epsilon\equiv\nabla\mathbf{v}$ under (a) Cartesian and (b) Protean coordinates. Dotted green lines correspond to the Viellefosse discriminant $Q^\star(R)=3/2^{2/3} |R|^{2/3}$.}\label{fig:QRpdfs}
\end{figure}

Figure~\ref{fig:QRpdfs} compares the joint distribution of the $Q$, $R$ invariants of the Cartesian $\boldsymbol\epsilon$ and Protean $\boldsymbol\epsilon^\prime$ velocity gradient tensors, where
\begin{align}
Q=-\frac{1}{2}\epsilon_{ij}\epsilon_{ji} && R\equiv\frac{1}{3}\epsilon_{ij}\epsilon_{jk}\epsilon_{ki},
\end{align}
and for the Protean velocity gradient tensor
\begin{align}
Q'=-\frac{1}{2}\epsilon'_{ij}\epsilon'_{ji}=-\epsilon^{\prime 2}_{11}-\epsilon^{\prime}_{11}\epsilon^{\prime}_{22}-\epsilon^{\prime 2}_{22} && R'\equiv\frac{1}{3}\epsilon'_{ij}\epsilon'_{jk}\epsilon'_{ki}=\epsilon^{\prime}_{11}\epsilon^{\prime}_{22}(\epsilon^{\prime}_{11}+\epsilon^{\prime}_{22}).
\end{align}
As such, all invariants of the Protean velocity gradient tensor fall below the Viellefosse discriminant~\citep{Meneveau:2011aa} 
\begin{equation}
Q^\star(R)=3/2^{2/3} |R|^{2/3},
\end{equation}
i.e., $Q^\prime<Q^\star(R^\prime)$, that demarcates spiral flow topologies ($Q>Q^\star$) from nodal topologies ($Q<Q^\star$). This is expected as the velocity gradient tensor is not objective (and spiral flow manifests as nodal flow in a rotating frame), hence the invariants $Q$, $R$ alter under the Protean transform which acts to remove material rotation by focussing on convergence to the Lyapunov vectors. The discrepancy between $Q$, $R$ and $Q'$, $R'$ highlights the need for objective measures when characterising Lagrangian quantities such as fluid deformation~\citep{Truesdell:1992aa} and classification of coherent flow structures~\citep{Haller:2015aa}.

\begin{figure}
\centering
\begin{tabular}{c c c}
\includegraphics[height=0.3\columnwidth]{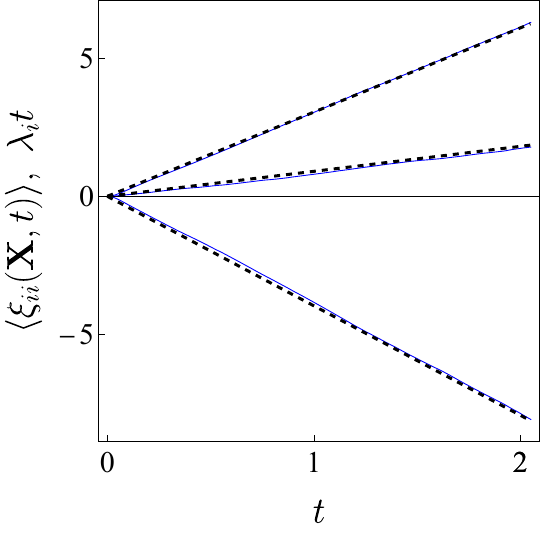}&
\includegraphics[height=0.3\columnwidth]{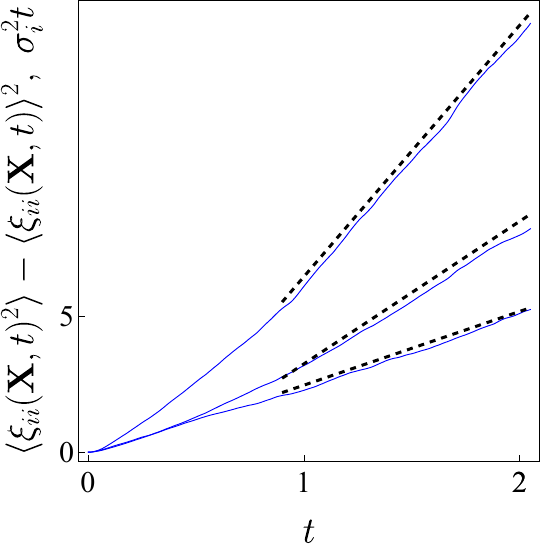}&
\includegraphics[height=0.3\columnwidth]{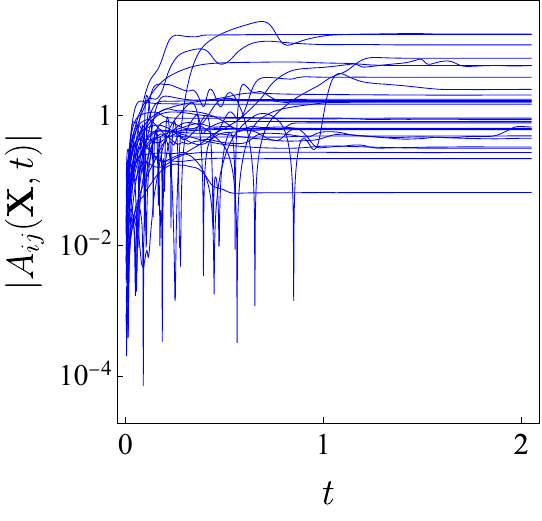}\\
(a) & (b) & (c)\\
\includegraphics[height=0.3\columnwidth]{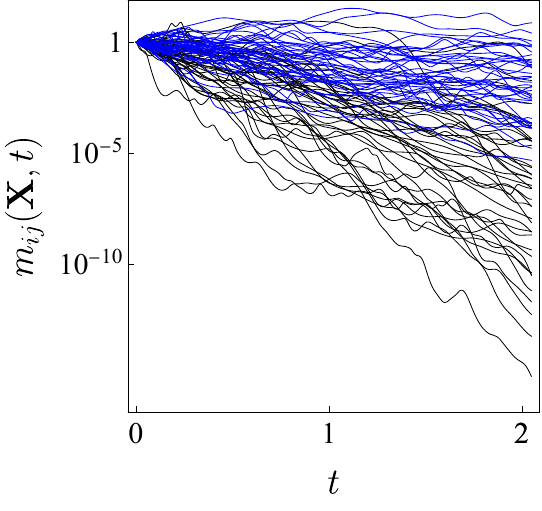}&
\includegraphics[height=0.3\columnwidth]{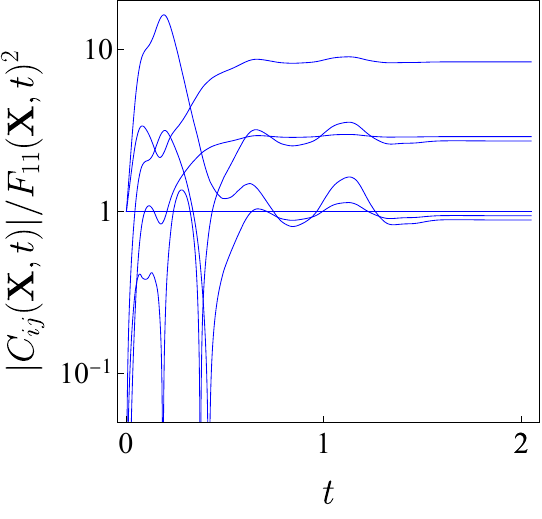}&
\includegraphics[height=0.3\columnwidth]{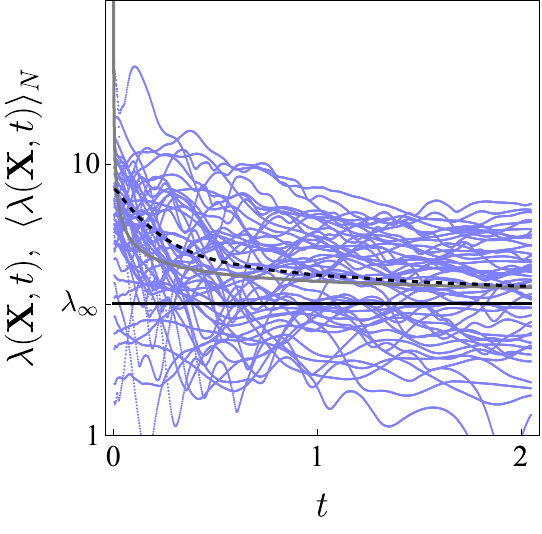}\\
(d) & (e) & (f)
\end{tabular}
\caption{(a) Evolution of (a) ensemble mean $\langle\xi_{ii}(t)\rangle_N$ and (b) ensemble variance $\langle\xi_{ii}(t)^2\rangle-\langle\xi_{ii}(t)\rangle_N^2$ of log-stretches (solid blue lines) over $10^3$ trajectories and respective analytic solutions $\lambda_{\infty,i} t$, $\sigma_{ii}^2 t$ (dashed black lines) for $i=1:3$. (c) Convergence of $A_{ij}(\mathbf{X},t)$ to steady value $a_{ij}(\mathbf{X})$ for 30 sample trajectories. (d) Decay of $m_{ij}(\mathbf{X},t)$ toward zero for 30 sample trajectories.
(e) Convergence of components $C_{ij}(\mathbf{X},t)/F_{11}(\mathbf{X},t)^2$ to a constant value for 30 sample trajectories. (f) Evolution of FTLE $\lambda(\mathbf{X},t)$ (light blue lines) over 30 sample trajectories and ensemble averaged FTLE $\langle\lambda(\mathbf{X},t)\rangle_N$ with Lagrangian time $t$ (black dashed line) toward Lyapunov exponent $\lambda_\infty$ (solid black line). The analytic expression (\ref{eqn:FTLEmean}) for $\langle\lambda(\mathbf{X},t)\rangle_N$ (solid gray line) is different to the numerical solution at short times as convergence to the central limit theorem is still developing.)}\label{fig:converge}
\end{figure}

Figure~\ref{fig:converge}(a), (b) show that the mean and variance of the log-stretches $\xi_{ii}(\mathbf{X},t)$ grow linearly in time (after some transient dynamics in the case of the latter) at rates that closely match the Lyapunov exponents $\lambda_{\infty,i}$ and variances $\sigma_{ii}^2$ respectively. This provides strong confirmation that the fluid deformation process is Fickian and validates the simple Brownian process (\ref{eqn:brownian}) for the evolution of $\xi_{ii}(\mathbf{X},t)$, Although the particle trajectories are only advected a few multiples $T/\tau_\lambda\approx 5$ of the stretching time $\tau_\lambda$, convergence of the terms $A_{ij}(\mathbf{X},t)$ to the steady values $a_{ij}(\mathbf{X})$ is readily apparent for most trajectories show in Figure~\ref{fig:converge}(c). Similarly, decay of the terms $m
_{ij}(\mathbf{X},t)$ with time for most trajectories is also apparent in Figure~\ref{fig:converge}(d). At longer times (not computed), all of the $A_{ij}$ and $m_{ij}$ terms will respectively converge to a constant and zero. Similarly, Figure~\ref{fig:converge}(e) shows convergence of all terms of the normalised Cauchy-Green tensor $\mathbf{C}(\mathbf{X},t)/F_{11}(\mathbf{X},t)^2$ to the constant values
\begin{equation}
\frac{\mathbf{C}(\mathbf{X},t)}{F_{11}(\mathbf{X},t)^2}
\rightarrow
\left(
\begin{array}{ccc}
1 & a_{12} & a_{13}\\
a_{12} & a_{12}^2 & a_{12}a_{13}\\
a_{13} & a_{12}a_{13} & a_{13}^2
\end{array}
\right).
\end{equation} 
This represents a significant simplification as for times $t\gg\tau_\lambda$, the Cauchy-Green tensor is simply a constant tensor scaled by $F_{11}^2$. Figure~\ref{fig:converge}(f) shows that the individual FTLEs $\lambda(\mathbf{X},t)$ fluctuate and slowly converge toward the leading Lyapunov exponent $\lambda_\infty$. After a short transient (associated with convergence to the CLT), the ensemble averaged FTLE $\langle\lambda(\mathbf{X},t)\rangle_N$ converges toward $\lambda_\infty$ as $1/\sqrt{t}$, in accordance with (\ref{eqn:FTLEmean}).

\section{Conclusions}
\label{sec:conclusions}

Fluid deformation controls many fluid-borne processes including 
solute mixing and dispersion, chemical and biological reactions, colloid transport and deposition and droplet breakup. Despite these widespread applications, several aspects of deformation are not well understood, particularly the link between fluid velocity (and gradients thereof) and deformation evolution.

This study presents an objective method to characterise the Lagrangian velocity gradient $\boldsymbol\epsilon$ and predict fluid deformation in 2D ($d=2$) and 3D ($d=3$) unsteady random flows. One barrier to such development has been that the Lagrangian velocity process in unsteady flows such as homogeneous isotropic turbulence (HIT) is non-Markovian in space and time, leading to strong intermittency~\citep{Meneveau:2011aa} and non-Fickian transport~\citep{Brandenburg:2004aa} that is difficult to characterise. However, recent studies~\citep{Dentz:2025aa} have shown that if these flows are space-time ergodic and non-separable in space and time, then they are rendered Markovian and Fickian on time scales longer than the Lagrangian temporal decorrelation scale $\tau_c$. For turbulent flows this corresponds to the Kolmogorov time scale $\tau_\eta$ for $\boldsymbol\epsilon$. Hence for $t\gg\tau_\eta$, the evolution of $\boldsymbol\epsilon$ can be described by a simple Brownian process along pathlines. 

A second barrier has been the lack of an objective frame in which to characterise $\boldsymbol\epsilon$. Transform of $\boldsymbol\epsilon$ into the rotating and translating \emph{Protean frame}~\citep{Adachi:1983aa,Lester:2018aa} corresponds to a continuous QR decomposition~\citep{Dieci:1997aa} that rapidly converges to the unique set of Lyapunov vectors $\mathbf{a}_i$~\citep{Froyland:2013aa} of the system. This transform renders the Protean velocity gradient tensor 
$\boldsymbol\epsilon^\prime$ objective in that ensemble averages of the diagonal components correspond to the Lyapunov spectra $\lambda_{\infty,i}$, $i=1:d$, and the off-diagonal components have zero mean and objectively characterise vorticity and shear. In this frame the deformation gradient tensor $\mathbf{F}^\prime$ is also upper triangular, leading to simple representations of the Cauchy-Green tensor $\mathbf{C}$ and associated finite-time Lyapunov exponents (FTLEs). 

In this study we use these approaches to develop a robust method to objectively characterise the velocity gradient in random unsteady flows and subsequently predict fluid deformation via a simple stochastic model. Application of this method to numerical simulations of a 2D random unsteady flow and DNS data of forced HIT shows that the evolution of $\boldsymbol\epsilon(t)$ is Fickian and Markovian for $t\gg\tau_c$. We also show that the deformation structure of these flows is quite simple in the Protean frame, consisting of strongly coupled (due to the divergence-free condition) diagonal components $\epsilon_{ii}^\prime$ (that objectively characterise stretching) that are weakly coupled to the off-diagonal  components $\epsilon_{ii}^\prime$ (that objectively characterise vorticity).  Based on an Ornstein–Uhlenbeck (OU) stochastic model for the Lagrangian evolution of $\boldsymbol\epsilon$, predictions of fluid deformation compare very well with direct numerical calculations, validating the underlying model assumptions and understanding of the deformation process.

This method provides a new approach to the problem of fluid deformation that underpins many fluid-borne processes. We anticipate that objective resolution of the Lagrangian velocity gradient tensor and associated strain develop will provide further insights into a wide range of random unsteady flows, from flows in random media to turbulence and beyond.

\appendix

\section{Equivalence of Protean and Material Strip Frames}
\label{app:strip}

Here we show that a coordinate frame that aligns with an infinitesimal 1D material strip is equivalent to the Protean frame. If we denote the coordinate frame that aligns with a material strip as $d\mathbf{x}^{\prime\prime}$, then in this frame $\mathbf{l}^{\prime\prime}(t)= d\mathbf{x}^{\prime\prime}=\rho(t)\hat{\mathbf{e}}_1^{\prime\prime}$ where $\rho(t)\equiv |\mathbf{l}(t)|/|\mathbf{l}(0)|$ is the elongation of the strip. Note that this frame shares many characteristics as the Protean frame, including rapid convergence of the direction of $\mathbf{l}(t)$ for different initial orientations $\mathbf{l}(0)$ due to stretching. In this frame $\mathbf{l}^{\prime\prime}(0)= d\mathbf{X}^{\prime\prime}=\hat{\mathbf{e}}_1^{\prime\prime}$, and so $\mathbf{l}(t)$ evolves as 
\begin{equation}
\mathbf{l}^{\prime\prime}(t)=\rho(t)\hat{\mathbf{e}}_1^{\prime\prime}=dx^{\prime\prime}=\mathbf{F}^{\prime\prime}\cdot d\mathbf{X}^{\prime\prime}=\mathbf{F}^{\prime\prime}\cdot\hat{\mathbf{e}}_1^{\prime\prime}    
\end{equation}
where $\mathbf{F}^{\prime\prime}$ is the deformation tensor in the material strip coordinate system. Hence $\hat{\mathbf{e}}_1^{\prime\prime}=(1,0,0)$ and $\rho(t)$ must be an eigenvector and eigenvalue of $\mathbf{F}^{\prime\prime}$, which can only occur if $\mathbf{F}^{\prime\prime}$ is upper triangular and $\rho(t)=F_{11}^{\prime\prime}(t)$. As $\mathbf{F}^{\prime\prime}$ may only be upper triangular if the associated velocity gradient tensor is so, $\boldsymbol\epsilon^{\prime\prime}$ is also upper triangular. As the continuous QR decomposition is unique for $t\gg\tau_{\Delta\lambda}$, then the material strip and Protean frames coincide after this initial transient, regardless of the initial line orientation. Hence the line stretch $\rho(t)$ evolves as
\begin{equation}
    \rho(t)=F_{11}^{\prime\prime}(t)=\exp\left(\int_0^t \epsilon^{\prime\prime}_{11}(t)\,\text{d}t\right)\approx F_{11}^{\prime}(t)=\exp\left(\int_0^t \epsilon^{\prime}_{11}(t)\,\text{d}t\right),
\end{equation}
and so the $F_{11}^\prime(t)$ in the Protean frame characterizes the stretching rate of material lines.

\section{2D Kraichnan Flow}
\label{app:Kraichnan}

The 2D time-dependent Kraichnan flow is generated according to
\begin{align}
    \mathbf v(\mathbf{x},t) = \sqrt{\frac{2}{N}} \sum\limits_{n=1}^N \cos(\mathbf k^{(n)} \cdot \mathbf{x} + \phi_n + \omega_n t) \mathbf p(\mathbf k^{(n)}),
\end{align}
where
\begin{align}
    p_1(\mathbf k) = \frac{k_2}{|\mathbf k|}, && p_2(\mathbf k) = \frac{-k_1}{|\mathbf k|}.
\end{align}
The $\mathbf k_j$ are sampled from a unit-Gaussian distribution. The phases $\phi_j$ are sampled from a uniform distribution between $0$ and $2 \pi$. The frequencies $\omega_j$ are sampled from a unit-Gaussian distribution. The flow field is divergence-free by construction. In the limit $N \to \infty$, $\mathbf v(\mathbf{x},t)$ is a multi-Gaussian random field with a Gaussian-shaped covariance function. Particle tracking is solved by an Euler-scheme,
\begin{align}
    \mathbf{x}(\mathbf{X},t + \Delta t) = \mathbf{x}(\mathbf{X},t) + \mathbf v[\mathbf{x}(\mathbf{X},t),t] \Delta t,\quad \mathbf{x}(\mathbf{X},t=0)=\mathbf{X},
\end{align}
with $\Delta t = 10^{-3}$. The components $\epsilon_{ij}(\mathbf{X},t)$ of the velocity gradient tensor $\boldsymbol \epsilon(\mathbf{X},t)$ are given by
\begin{align}
    \epsilon_{ij}(\mathbf{X},t) &= (-1)^i \sqrt{\frac{2}{N}} \sum\limits_{n=1}^N \cos[\mathbf k^{(n)} \cdot \mathbf{x}(\mathbf{X},t) + \phi_n + \omega_n t] p_i(\mathbf k^{(n)}) k^{(n)}_{j}.
\end{align}

\section{Evolution Equations for Orientation Angles $\alpha_i(t)$}
\label{app:alpha}

The evolution equations for $\alpha_i(t)$ are generated by first constructing the rotation matrix $\mathbf{Q}(\mathbf{X},t)=\mathbf{Q}_3(\alpha_3)\mathbf{Q}_2(\alpha_2)\mathbf{Q}_1(\alpha_1)$ via (\ref{eqn:Qi}). Insertion of $\mathbf{Q}$ into (\ref{eqn:rotn}) and setting $\epsilon^\prime_{ij}=0$ for $j>i$ yields the following evolution equations

\begin{equation}
    \begin{split}
&\alpha_1'(t)=\cos ^2\alpha_1 (\sin \alpha_2 ((\epsilon_{22}-\epsilon_{11}) \sin (2\alpha_3)+(\epsilon_{12}+\epsilon_{21}) \cos (2\alpha_3))\\
&+\cos \alpha_2 ((\epsilon_{23}+\epsilon_{32}) \cos \alpha_3-(\epsilon_{13}+\epsilon_{31}) \sin\alpha_3))\\
&+\frac{1}{4} \cos\alpha_3 ((\epsilon_{13}+\epsilon_{31}) \sin (2\alpha_1) \sin (2\alpha_2)-4 \epsilon_{23} \sec\alpha_2)+\epsilon_{13} \sec \alpha_2 \sin \alpha_3\\
&+\frac{1}{8} \sin (2\alpha_1) \Big[-(\epsilon_{11}-\epsilon_{22}) (\cos (2\alpha_2)-3) \cos (2\alpha_3)-(\epsilon_{12}+\epsilon_{21}) (\cos (2\alpha_2)-3) \sin (2\alpha_3)\\
&+2 (\epsilon_{23}+\epsilon_{32}) \sin (2\alpha_2) \sin \alpha_3-2 \cos ^2\alpha_2 (\epsilon_{11}+\epsilon_{22}-2 \epsilon_{33})\Big]
    \end{split}
\end{equation}

\begin{equation}
    \begin{split}
&\alpha_2'(t)=\frac{1}{8} (-2 \sin (2\alpha_2) ((\epsilon_{11}-\epsilon_{22}) \cos (2\alpha_3)+(\epsilon_{12}+\epsilon_{21}) \sin (2\alpha_3)+\epsilon_{11}+\epsilon_{22}-2 \epsilon_{33})\\
&-4 \cos \alpha_3 ((\epsilon_{13}+\epsilon_{31}) \cos (2\alpha_2)-\epsilon_{13}+\epsilon_{31})-4 \sin \alpha_3 ((\epsilon_{23}+\epsilon_{32}) \cos (2\alpha_2)-\epsilon_{23}+\epsilon_{32}))
    \end{split}
\end{equation}

\begin{equation}
    \begin{split}
        \alpha_3'(t)=&\frac{1}{2} \Big[2 \tan \alpha_2 (\epsilon_{13} \sin \alpha_3-\epsilon_{23} \cos \alpha_3)+(\epsilon_{22}-\epsilon_{11}) \sin(2\alpha_3)+(\epsilon_{12}+\epsilon_{21}) \cos (2\alpha_3)\\
        &-\epsilon_{12}+\epsilon_{21}\Big]
    \end{split}
\end{equation}

\bibliographystyle{jfm.bst}
\bibliography{reflist}

\backsection[Acknowledgments]{Funded by the European Union under the grant ERC XXX. Views and opinions expressed are however those of the author(s) only and do not necessarily reflect those of the European Union or the European Research Council Executive Agency. Neither the European Union nor the granting authority can be held responsible for them.}

\backsection[Funding]{MSCA COFUND REDI, 101034328}

\backsection[Declaration of interests]{The authors report no conflict of interest.}

\backsection[Data availability statement]{The numerical codes and relevant data will be made available upon publication}

\backsection[Author ORCIDs]{D. Lester, https://orcid.org/0000-0003-2927-1384}

\backsection[Author contributions]{D. R. Lester and M. Dentz devised the conceptual approach and developed theory. All authors contributed to revising the manuscript.}

\end{document}